\documentclass[aip,jcp,numerical,reprint]{revtex4-1}

\usepackage{dcolumn,graphicx,bm}
\def\bra#1{\mathinner{\langle{#1}|}}
\def\ket#1{\mathinner{|{#1}\rangle}}

\def\Bra#1{\left<#1\right|}
\def\Ket#1{\left|#1\right>}
\newcommand{\tr}{{\rm tr}}
\usepackage{epstopdf}
\usepackage{amsmath,amstext,amssymb,graphicx,bm,dcolumn,color,url}

\begin{document}

\title{Time dynamics of multiparty quantum correlations indicate energy transfer route in light-harvesting complexes}

\author{Titas Chanda}
\affiliation{Harish-Chandra Research Institute, Chhatnag Road, Jhunsi, Allahabad 211 019, India}
\author{Utkarsh Mishra}
\affiliation{Harish-Chandra Research Institute, Chhatnag Road, Jhunsi, Allahabad 211 019, India}
\author{Aditi Sen(De)}
\affiliation{Harish-Chandra Research Institute, Chhatnag Road, Jhunsi, Allahabad 211 019, India}
\author{Ujjwal Sen}
\affiliation{Harish-Chandra Research Institute, Chhatnag Road, Jhunsi, Allahabad 211 019, India}

\date{\today}

\begin{abstract}
The time-evolution of multiparty quantum correlations as quantified by monogamy scores and bipartition collections of quantum correlations is investigated for light-harvesting complexes modeled by the fully connected and the Fenna-Mathews-Olson (FMO) networks. The dynamics consists of a coherent term as well as dissipative, dephasing, and sink operator terms. The multiparty quantum correlation reveals important information regarding the sharability of quantum correlations in the networks, which allow us to 	categorize the network sites into three distinct groups in the FMO complex and to predict the structural geometry of the complex. In particular, we show that the relative values of the ingredients of  multiparty quantum correlation measures in the time dynamics clearly indicate the primary route of energy transfer from the antenna to the bacterial reaction center in the FMO complex.

\end{abstract}

\maketitle
\section{Introduction}
Quantum correlations \cite{HHHHRMP,MODIRMP} are known to play a crucial role in a wide variety of physical phenomena in ultracold gas, solid state, nuclear magnetic resonance, and other systems \cite{LewesnteinAdp, FazioRMP,  others-theory+exp}. In the last few years, quantum coherence and quantum correlations have been claimed to be of relevance in certain biological processes \cite{Naturephyswhaley, NatPhysNori,SR}.  
Two important biological phenomena in which quantumness may play a role include avian magnetoreception, used by  migratory birds  for efficient navigation \cite{ritzbio}, and the Fenna-Mathews-Olson (FMO) light-harvesting protein complex of green sulfur bacteria, responsible for photosynthesis \cite{Naturephyswhaley, NatPhysNori,pnas,engel,collini,Renger_biop}. In the latter process,  the FMO complex plays the role of the mediator to transfer excitation energy from the light-harvesting chlorosome antennae to a reaction center. Recently,  it was argued that the efficient transfer of energy in photosynthesis can not be explained by the classical incoherent hopping model \cite{caruso_2009,SLloyd}.  On the other hand, several studies show that quantum coherence is essential in excitation energy transfer in the FMO complex \cite{Wilde,SLloyd,castro_2008_prb, Castro_njp,caruso_2009,caruso_2010_pra,Naturephyswhaley,chin_njp_2010,semiao_njp_2010,Thilagam, Naturephyswhaley, NatPhysNori,ekert}. Specifically, the dynamics of entanglement  
under the influence of dissipative environments, both Markovian and non-Markovian, have been extensively investigated \cite{caruso_2010_pra,rupak_jpc,Castro_njp}. 

Most of the studies in this direction are restricted to bipartite quantum correlation measures (for exceptions, see \cite{Love, Thilagam}). However, there exist several phenomena in the domain of quantum information and many-body physics
which can not signaled and explained by bipartite quantum correlation measures, while multisite quantum correlations provide an adequate 
description (see e.g. \cite{manab,secret_sharing,cluster_state, steane}).
Also, recent experimental breakthroughs have  ensured that multiparty quantum states can be created and their multiparty quantum correlations can be detected \cite{multiexp}.

An obstacle in the study of multipartite quantum correlations is the usual unavailability of computable multipartite measures.  
One avenue to overcome this difficulty is to work with the class of multipartite quantum correlation measures based on the concept of monogamy.    
Qualitatively, monogamy of a quantum correlation measure says that among three parties sharing a quantum state, if two are highly quantum correlated, then the third party can only possess a negligible amount of quantum correlation individually with the other two.  Note that classical correlations do not satisfy any such monogamy condition. This qualitative concept of monogamy can and has been quantified \cite{wootters,seealso}, and  leads to multiparty quantum correlation measures, referred to as monogamy scores \cite{wootters,our_gang}.

Another strategy to understand multiparty quantum correlations is to look at the collection of the bipartite quantum correlations across different partitions of the system. This is akin to the concept of entanglement entropy and area law \cite{area_law}, where the scaling of entanglement of a part of a many-body system to the rest is used to understand the cooperative phenomena in the system. 

In this paper, we investigate the dynamics of monogamy-based multipartite quantum correlation measures, specifically, negativity and discord monogamy scores,  as well as collections of quantum correlations in the different bipartitions. The investigations are carried out for the fully connected network \cite{ekert,caruso_2009} as well as for the FMO complex. To understand the effects of decoherence in the energy transfer, the evolution of multiparty quantum correlations are studied through the Lindblad mechanism, including dissipation and dephasing effects, with the initial states being close to the antenna. 
We find that the behavior of multipartite quantum correlation measures depend both on the initial state as well as the ``nodal'' observer used in the monogamy scores.  The results show that in the FMO complex, the negativity monogamy score is more robust against noise than its bipartite counterpart, irrespective of the choice of the nodal observer and the initial state. Recent findings show that the FMO complex is made of seven inequivalent chromophore sites and a sink \cite{caruso_2010_pra}. 
We  observe a complementary behavior between the dynamics of negativity  and discord monogamy scores in the presence of noisy environments, both in the fully connected network model and in the FMO complex. In particular, we find that in the FMO complex, the modulus of the discord monogamy score possesses much smaller values as compared to the quantum discord present in the constituent bipartite states -- the situation is opposite in the case of negativity. The findings clearly indicate that to detect quantum correlations in the FMO complex, multipartite entanglement measures are  more effective than the bipartite ones. The opposite is true for quantum discord. Moreover the dynamics of quantum correlations allows us to classify the seven inequivalent sites of the FMO complex into three groups, which in turn helps us to predict the structural arrangement of the complex. Interestingly, if we look at collections of the quantum correlations in different bipartitions, and investigate their behavior as they evolve in time, then their relative values indicate the route through which the energy is transformed from the antenna to the bacterial reaction center, as has been estimated in recent observations \cite{Love, Naturephyswhaley, chin_njp_2010}. 

The paper is organized as follows. In Sec. \ref{sec:network}, we discuss the network models for light-harvesting complexes
and describe their time evolution that involves dissipative and dephasing
effects. In Sections \ref{sec:qcmeasures} and \ref{sec:mngscore}, we present the bipartite and multipartite
quantum correlations used in this paper. The results regarding the behavior of the monogamy
scores and bipartite quantum correlations with time are presented in
Sec. \ref{sec:mngfmofn}. In particular, Sec. \ref{sec:mngfcn} presents the results for fully connected network model, while those for FMO complex are discussed in Sec. \ref{sec:mngfmo}. 
In Sec. \ref{sec:energytransfer}, we show that ingredients of  multiparty quantum correlation measures can detect the primary route of energy transfer from the antenna to the bacterial reaction center in the FMO complex. We conclude in Sec. \ref{sec:conclusion}.

\section{The network model for light-harvesting complexes}
\label{sec:network}
In this section, we  review important abstract network models for quantum transport, where the local sites undergo both local dissipation and dephasing noise, and the excitation transfer to a reaction center is designed via an irreversible coupling to a preferred  trapping site \cite{Renger_biop, caruso_2009, Naturephyswhaley, ekert, enk}. This model is a basic framework for light-harvesting complexes as they are typically constituted of multiple chromophores which irreversibly transfer excitations to the reaction center. In order to illuminate the basic phenomena clearly, it is usual to consider that the relevant complexes are composed of several distinct two-level sites. In a network of $N$ sites, an excitation in the site $j$ is described as
\begin{eqnarray}
\ket{j} =\left( \ket{e}_{j} \bigotimes_{\substack{i=1 \\ i \neq j }}^{N} \ket{g}_{i}\right) \otimes \ket{g}_{N+1},
\label{site_state}
\end{eqnarray}
where $\arrowvert g\rangle$ denotes the absence of an excitation and $\arrowvert e\rangle$ represents the presence of an excitation at a particular site. The site $N+1$ is also included, to be treated as a ``\textit{sink}" site. A sink state $\ket{N+1}$, indicating that the exciton is trapped to the reaction center, is given as follows:
\begin{eqnarray}
\ket{N+1} = \left(\bigotimes_{i=1}^{N} \ket{g}_{i}\right) \otimes \ket{e}_{N+1}.
\label{sink_state}
\end{eqnarray}
Similarly the ground state $\ket{0}$, that represents the loss of the exciton, is given by
\begin{eqnarray}
\ket{0} = \bigotimes_{i=1}^{N+1}\ket{g}_{i}.
\label{ground_state}
\end{eqnarray}
The density matrix which characterizes the quantum state of the whole network is 
\begin{eqnarray}
\rho = \sum_{i,j \in \{0,1,...,N+1\}} \rho_{ij} \ket{i}\bra{j}.
\label{density}
\end{eqnarray}
The coherent exchange of excitations between sites in the network is governed by a simple Hamiltonian dynamics.  The dephasing and the dissipation caused by the environment are modeled using local Lindblad terms. The Hamiltonian for the coherent evolution of a network of $N$ sites is given by
\begin{eqnarray}
H = \sum_{j=1}^{N} \hbar \omega_{j} \sigma_{j}^{+} \sigma_{j}^{-} + \sum_{\substack{i,j=1 \\ i \neq j}}^N \hbar v_{ij} (\sigma_{i}^{+} \sigma_{j}^{-} + \sigma_{j}^{+} \sigma_{i}^{-}), 
\label{eq:NM_hamil}
\end{eqnarray}
where $\sigma_{j}^{+}$ and $\sigma_{j}^{-}$ are the raising and lowering operators respectively for the site $j$, and have the form $\sigma_{j}^{+} = \ket{j}\bra{0}$ and  $\sigma_{j}^{-}=\ket{0}\bra{j}$. Also, $\hbar \omega_j$ denotes the on-site excitation energy at $j$ whereas $\hbar v_{ij}$ represents the coupling energy between the sites $i$ and $j$. We assume that the system  is affected by two distinct types of noise due to environmental effects: (i) dissipation of the exciton that transfers the excitation energy of site $j$ to the environment, and (ii) a dephasing interaction with the environment that destroys the phase coherence of the system. Both types of noise processes can be described using a Markovian master equation with local dephasing and dissipation terms by the following Lindblad super-operators:
\begin{eqnarray}
\mathcal{L}_{diss}(\rho) &=& \sum_{j=1}^{N} \Gamma_j \left[ 2\sigma_{j}^{-}\rho\sigma_{j}^{+} - \{\sigma_{j}^{+}\sigma_{j}^{-},\rho\}  \right], \\
\mathcal{L}_{deph}(\rho) &=& \sum_{j=1}^{N} \gamma_j \left[ 2\sigma_{j}^{+}\sigma_{j}^{-}\rho\sigma_{j}^{+}\sigma_{j}^{-} - \{\sigma_{j}^{+}\sigma_{j}^{-},\rho\}  \right],
\label{lind_diss_deph}
\end{eqnarray}
where $\Gamma_{j}$ and $\gamma_{j}$ denote respectively the dissipation and dephasing rates of the noise processes for site $j$. The trapping of the exciton in the reaction center by an irreversible decay process from a ``preferred'' site $k$ is described by the Lindblad super-operator,

\begin{eqnarray}
\mathcal{L}_{sink}(\rho) = \Gamma_{N+1}[2\sigma_{N+1}^{+}\sigma_{k}^{-}\rho\sigma_{k}^{+}\sigma_{N+1}^{-} \nonumber \\
-\{\sigma_{k}^{+}\sigma_{N+1}^{-}\sigma_{N+1}^{+}\sigma_{k}^{-},\rho\} ].
\end{eqnarray} 
The evolution of the density operator $\rho$, given in Eq. (\ref{density}), is governed by the equation
\begin{eqnarray}
\label{eq:evolvedm}
\dot{\rho} = - i [H,\rho] + \mathcal{L}_{diss}+ \mathcal{L}_{deph}+ \mathcal{L}_{sink}.
\end{eqnarray}
The population at site $j$ at time $t$ is obtained by $p_{j}(t)=\bra{j}\rho(t)\ket{j}$, whereas the population transferred to the reaction center is given by $p_{sink}(t) = 2 \Gamma_{N+1} \int_{0}^{t} p_{k}(t')\text{d}t'$.

\begin{figure*}
\includegraphics[width=0.45\linewidth]{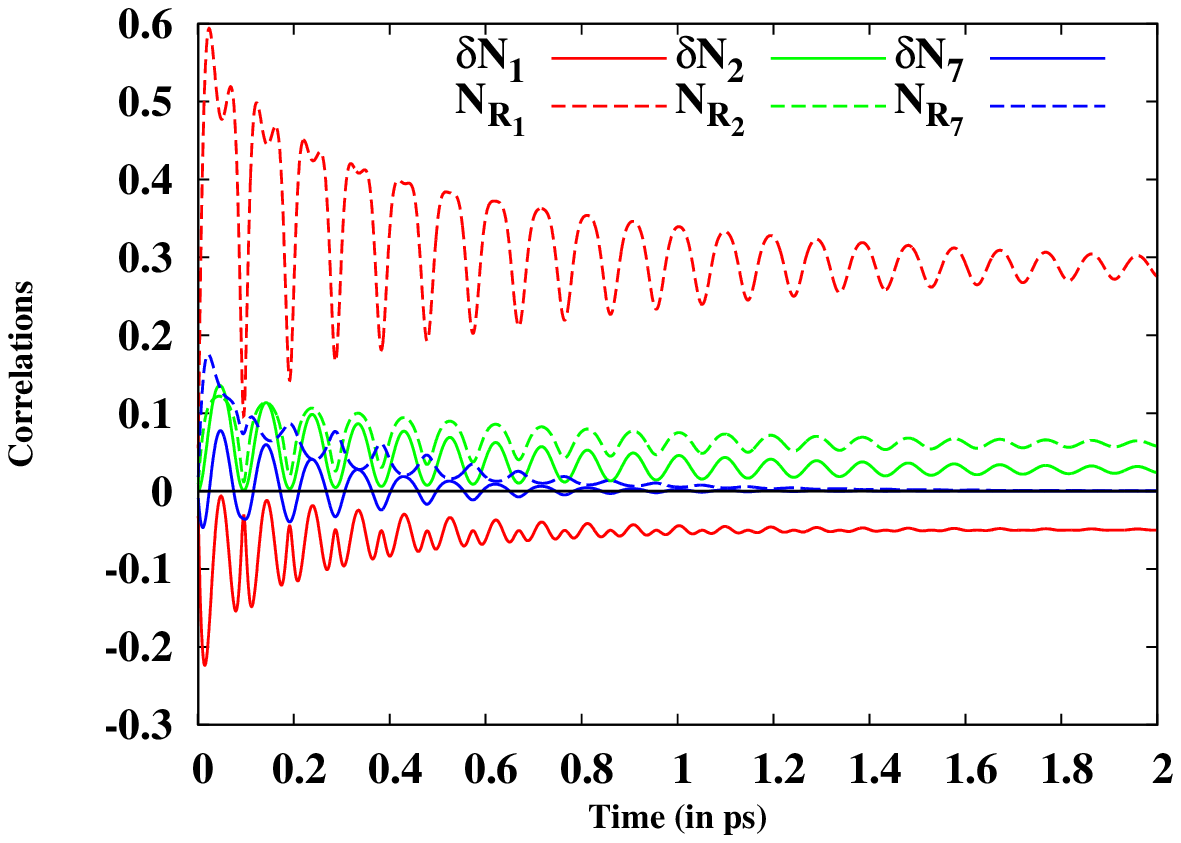}
\includegraphics[width=0.45\linewidth]{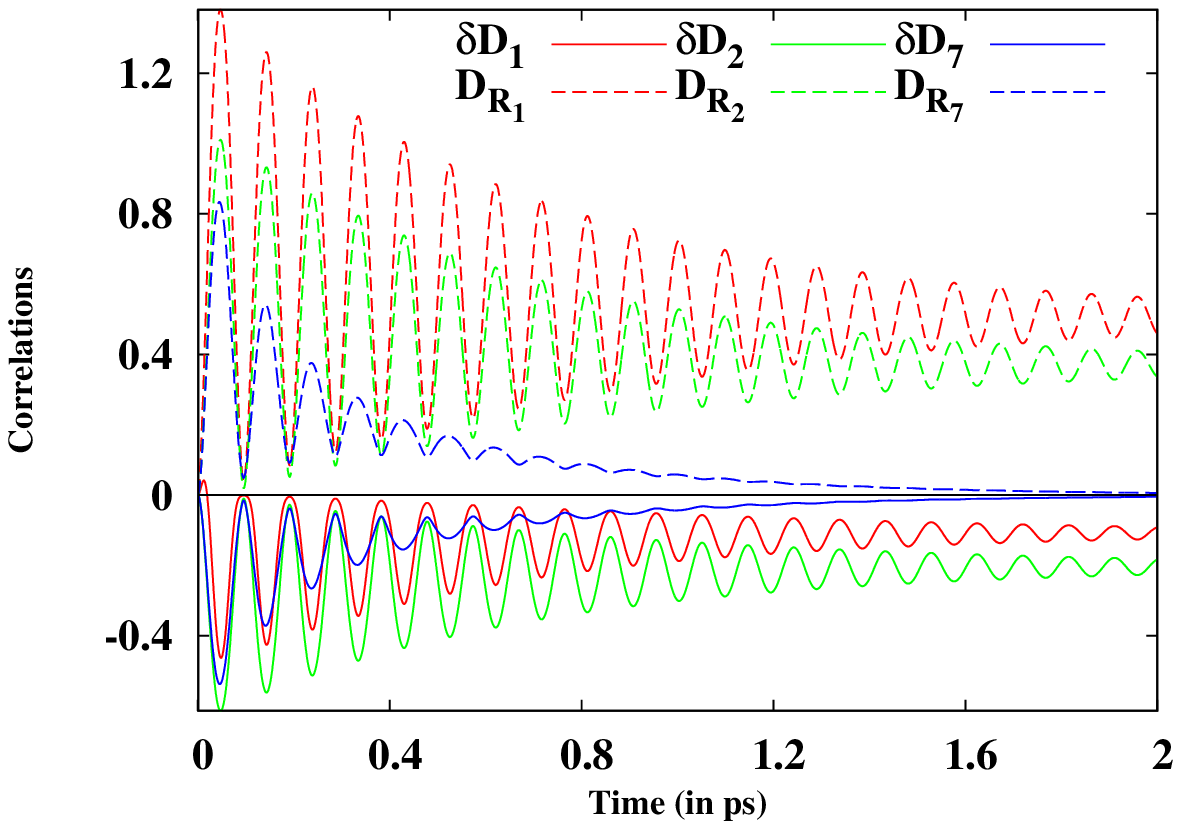}
\caption{(Color online.) Monogamy scores in the FCN in the clean case. Monogamy scores for negativity ($\delta  N$) and for discord ($\delta  D$) are plotted as ordinates against time along as abscissae for the seven-site FCN network in the clean case. The different colors are for monogamy scores with respect to different nodal observers. The solid lines represent the monogamy scores while the dashed lines are for the bipartite contributions. The quantities plotted as the ordinates in the left panel are in ebits, while those in the right one are in bits.}
\label{fig:msfcn_no}
\end{figure*}

\begin{figure}
\includegraphics[width=\linewidth]{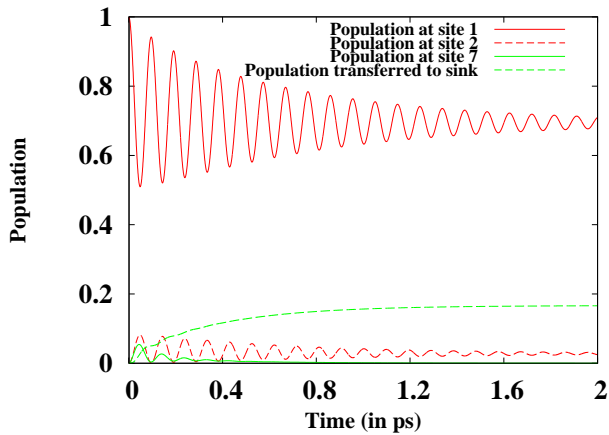}
\caption{(Color online.) Population at different sites as functions of time for the FCN in the clean case. The ordinates are dimensionless.}
\label{pop_FCN}
\end{figure}

\subsection{Fully Connected Network Model}
\label{subsec:fcn}
The fully connected network (FCN) model is a simple network model which helps us to understand the effects of various environmental effects on quantum transport. Thus, before going over to the more complex FMO dynamics, it is helpful to study this simpler network model. The FCN is an abstract model in which all the coupling constants in the Hamiltonian (Eq. (\ref{eq:NM_hamil})) are equal, i.e., $\hbar v_{ij} = J$ for all $i \neq j$. In case of ``uniform'' FCN, i.e., when  $\omega_j$, $\Gamma_j$, and $\gamma_j$ are equal for all sites, an exact analytical solution of the density matrix can be found for arbitrary (positive) integral values of $N$ \cite{caruso_2009}. This exact solution provides useful insights into various mechanisms that
contribute to the dephasing-assisted transport as well as to the quantum correlations involved.

\subsection{FMO Complex}
\label{subsec:fmo}

The FMO complex is a pigment-protein complex which is believed as the main contributor to the ultra-efficient energy transfer from the light-harvesting chlorosomes to the bacterial reaction center in green sulfur bacteria. It is a trimer of three identical units, each composed
of seven bacteriochlorophyll  \textit{a} molecules lodged in a scaffolding of protein molecules. Generally, it is modeled as a connected network of \textit{seven} chromophore sites corresponding to \textit{seven} bacteriochlorophyll \textit{a} molecules with site-dependent coupling strengths and site energies.
The matrix form of the Hamiltonian (in the site basis $\{\ket{j}\}_{j=1}^7$) responsible for the coherent dynamics of the complex is as follows \cite{Renger_biop}:
\small
\begin{equation}
H = \begin{pmatrix}
215 & -104.1 & 5.1 & -4.3 & 4.7 & -15.1 &-7.8 \\
-104.1 & 220 & 32.6 & 7.1 & 5.4 & 8.3 & 0.8 \\
5.1 & 32.6 & 0 & -46.8 & 1.0 & -8.1 & 5.1 \\
-4.3 & 7.1 & -46.8 & 125 & -70.7 & -14.7 & -61.5\\
4.7 & 5.4 & 1.0 & -70.7 & 450 & 89.7 & -2.5\\
-15.1 & 8.3 & -8.1 & -14.7 & 89.7 & 330 & 32.7 \\
-7.8 & 0.8 & 5.1 & -61.5 & -2.5 & 32.7 & 280 
\end{pmatrix},
\label{FMO_hamil}
\end{equation}
\normalsize

\noindent where the numbers are given in units of $\text{cm}^{-1}$, a general convention in spectroscopic experiments. The incoherent part of the dynamics is the same as the network model described above. Recent work suggests that site $1$ and $6$ are closest to the  chlorosome antenna and are thus most likely to be the initial state of the FMO complex in the dynamics \cite{SLloyd}, whereas  site $3$ is the preferred site, which is coupled to the reaction center at site $8$ (sink) \cite{Renger_biop}. We choose the trapping rate, as in \cite{caruso_2009, Wilde}, to be $\Gamma_{8} = 62.8/1.88 \ \text{cm}^{-1}$, corresponding to about $6.3 \ \text{ps}^{-1}$. We also assume that the dissipation rates are same for all sites and have the value $\Gamma_{j} = \Gamma_{diss} = 1/(2 \times 188) \ \text{cm}^{-1}$, corresponding to about $5 \times 10^{-4} \ \text{ps}^{-1}$. Finally, we set the optimized dephasing rates  $\gamma_{j}$ to be $\{0.157,9.432,7.797,9.432,7.797,0.922,9.433\} \ \text{ps}^{-1}$, for time $t=5$ ps, as in Ref. \cite{caruso_2009}.

\begin{figure*}[ht]
\includegraphics[width=0.45\linewidth]{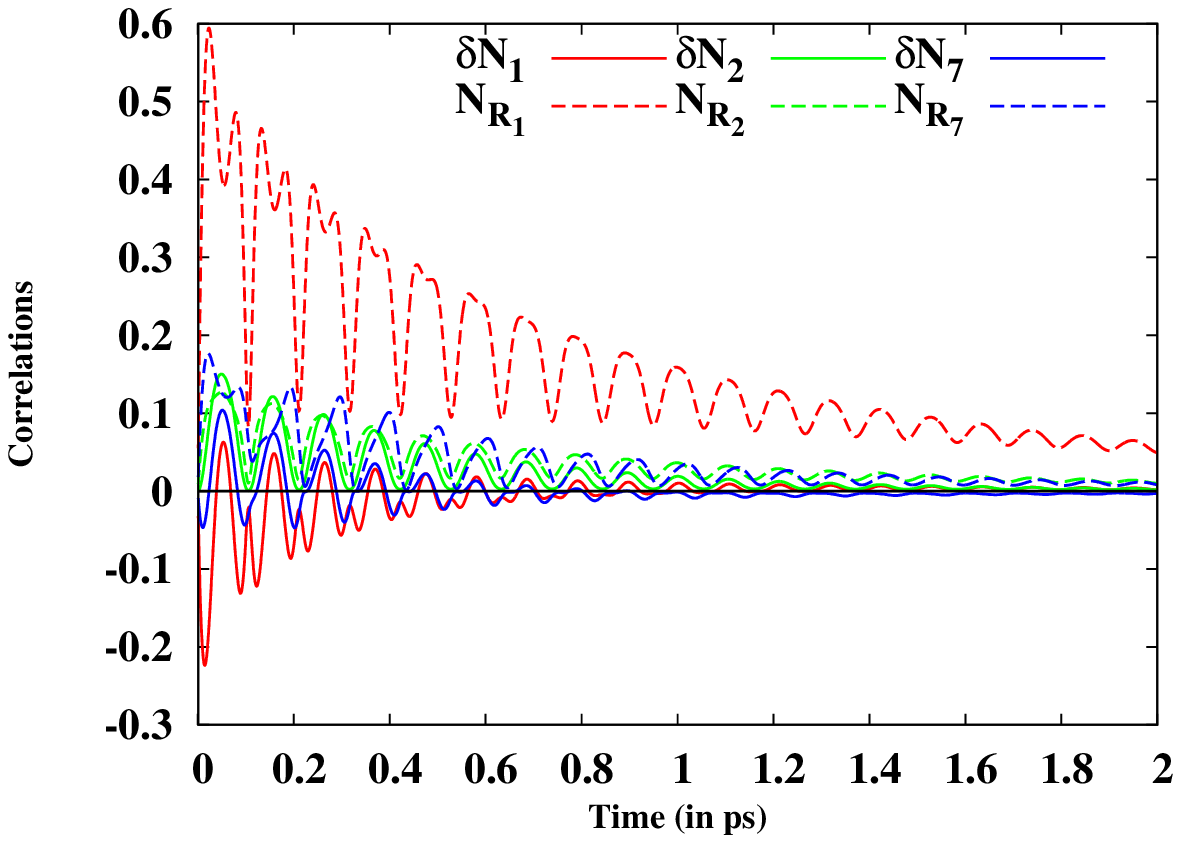}
\includegraphics[width=0.45\linewidth]{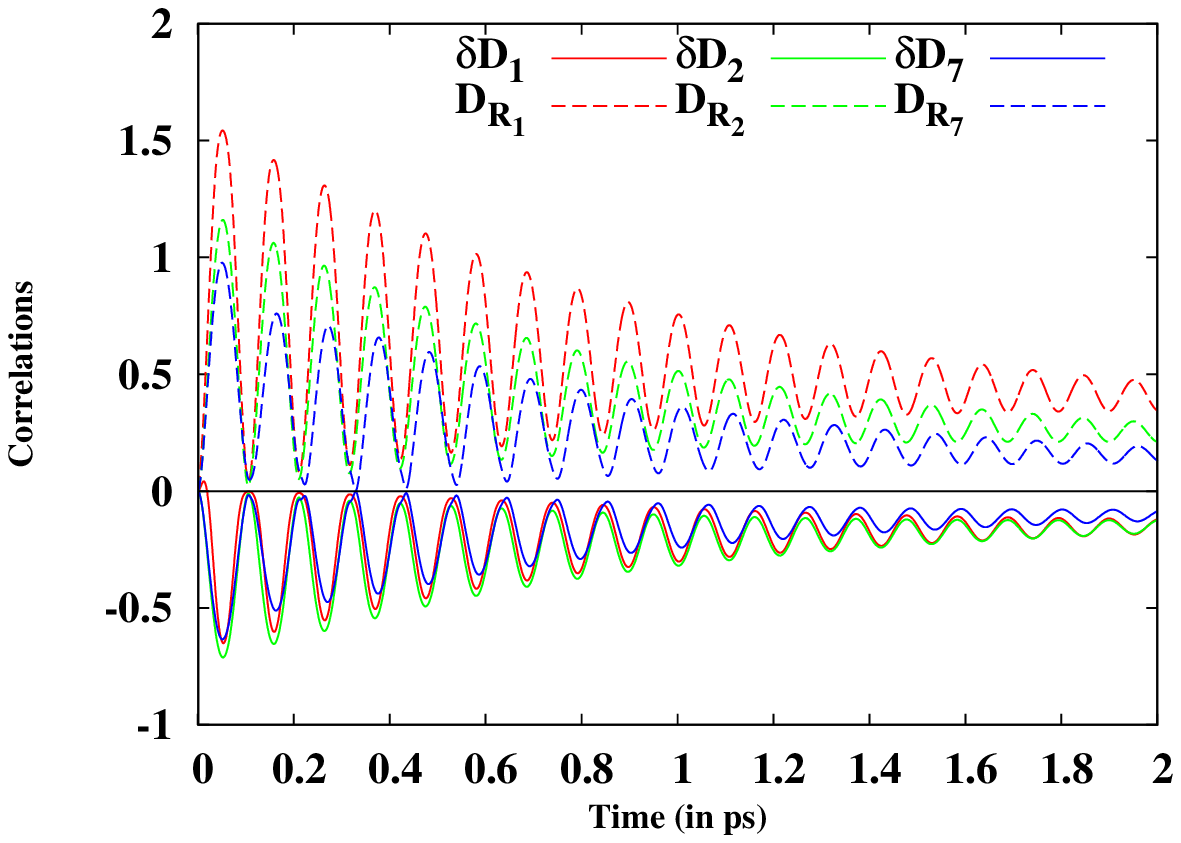}
\caption{(Color online.) Monogamy scores in the FCN in the case of energy mismatch. See Sec. \ref{subsec:emis} for the values of the parameters. All other descriptions remain the same as in Fig. \ref{fig:msfcn_no}.}
\label{fig:msfcn_en}
\end{figure*}

\section{Bipartite Quantum Correlations}
\label{sec:qcmeasures}
Quantum correlations, in the form of entanglement \cite{HHHHRMP} and quantum discord \cite{MODIRMP}, considered to be unique characteristics of quantum systems, are useful resources for many quantum information and computational tasks \cite{GISINRMP}. Among entanglement measures, we will mainly focus on the negativity \cite{werner-vidal}, which is based on the Peres-Horodecki criterion of separability \cite{perespt}. Among other things, negativity has the advantage as it is computable for arbitrary states in arbitrary bipartite dimensions. Quantum discord, is an information theoretic measure of quantum correlations and is in some sense a more fine-grained detector of quantum correlations as compared to entanglement. The measures are however equivalent, though not necessarily of equal numerical value, for pure bipartite quantum states.

For a bipartite quantum state, $\rho_{AB}$, the negativity is defined as the absolute sum of the negative eigenvalues of the partial transposed state: 
\begin{equation}
\label{eq:neg}
N_{A:B}\equiv N(\rho_{A:B})=\frac{||\rho_{AB}^{T_{A}}||-1}{2},
\end{equation}
where $||A|| = \mbox{Tr}\sqrt{A^{\dagger}A}$ is the trace norm of the matrix $A$ and 
$\rho_{AB}^{T_{A}}$ is the partially transposed state with partial transposition \cite{perespt} being taken with respect to party $A$. The value of the negativity does not depend on which party the partial transposition is performed.  

Quantum discord of the bipartite quantum state, $\rho_{AB}$, is defined in terms of the quantum mutual information, ${\cal I}$  and ``classical correlation'', ${\cal J}$, as  \cite{zurek,henderson-vedral}
 \begin{equation}
D_{A:B} \equiv D(\rho_{A:B})={\cal I}(\rho_{AB}) - {\cal J}(\rho_{AB}).
\end{equation}
Here, ${\cal I}(\rho_{AB})$ is defined as
 \begin{equation}
{\cal I}(\rho_{AB})=S(\rho_{AB})-S(\rho_{A})-S(\rho_{B}),
 \end{equation}
 and interpreted as the total correlation present in the quantum state $\rho$ \cite{henderson-vedral}, where $S(\varrho) = \tr(\varrho\log_2 \varrho)$ is the von Neumann entropy of $\varrho$. The quantity  ${\cal J}$ is defined as 
 \begin{equation}
 {\cal J}=S(\rho_{B})-S(\rho_{B|A}),
 \end{equation}
and interpreted as the classical correlations in $\rho$, where
\begin{equation}
S(\rho_{B|A})=\mbox{min}\sum_{i}p_{i}S(\rho_{B|i})
\end{equation}
is the minimal average conditional entropy obtained by performing measurement on the $A$ part and averaging over all measurement outcomes with the corresponding outcome probabilities being  $p_{i}$. The minimization is over all projection measurements on the $A$ part. For nonsymetric states, the value of quantum discord  depends on the part on which the measurement is performed. We will always be considering situations where measurement is performed on the first system. 

\begin{figure*}
\includegraphics[width=0.45\linewidth]{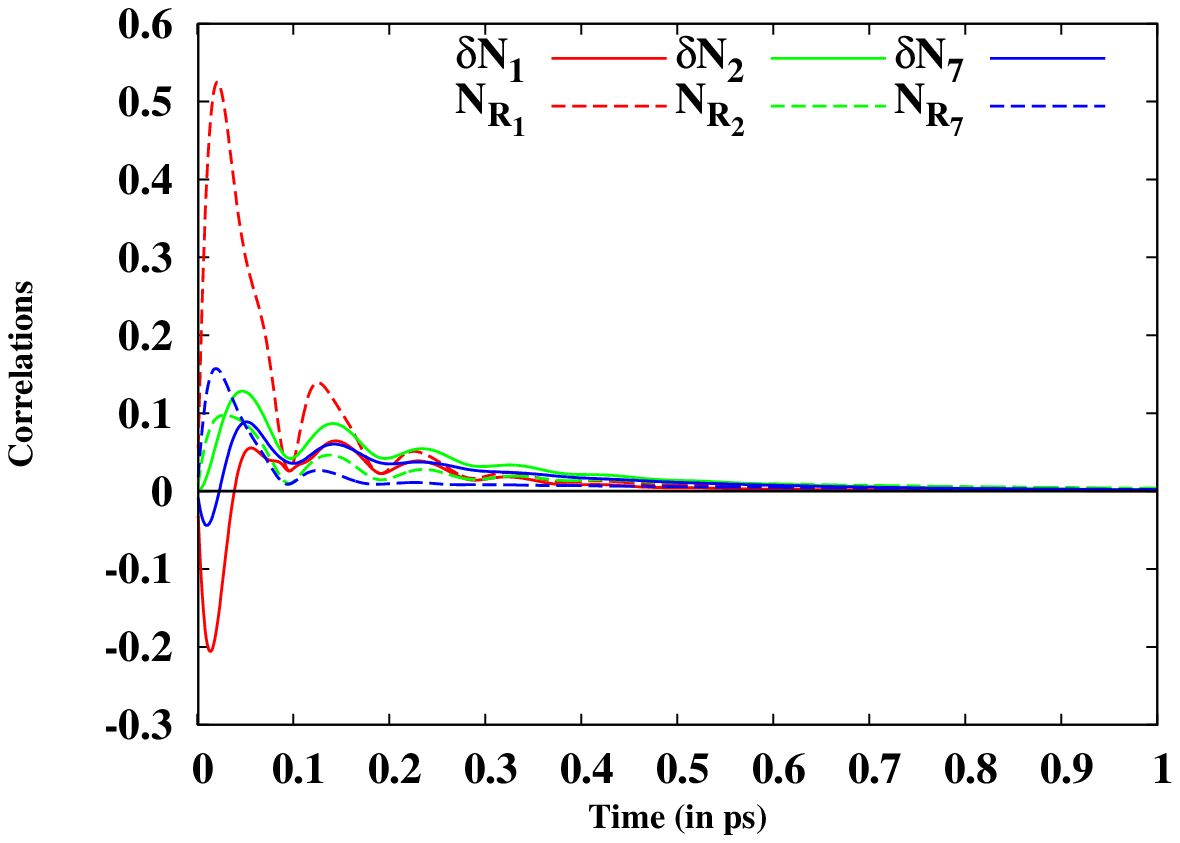}
\includegraphics[width=0.45\linewidth]{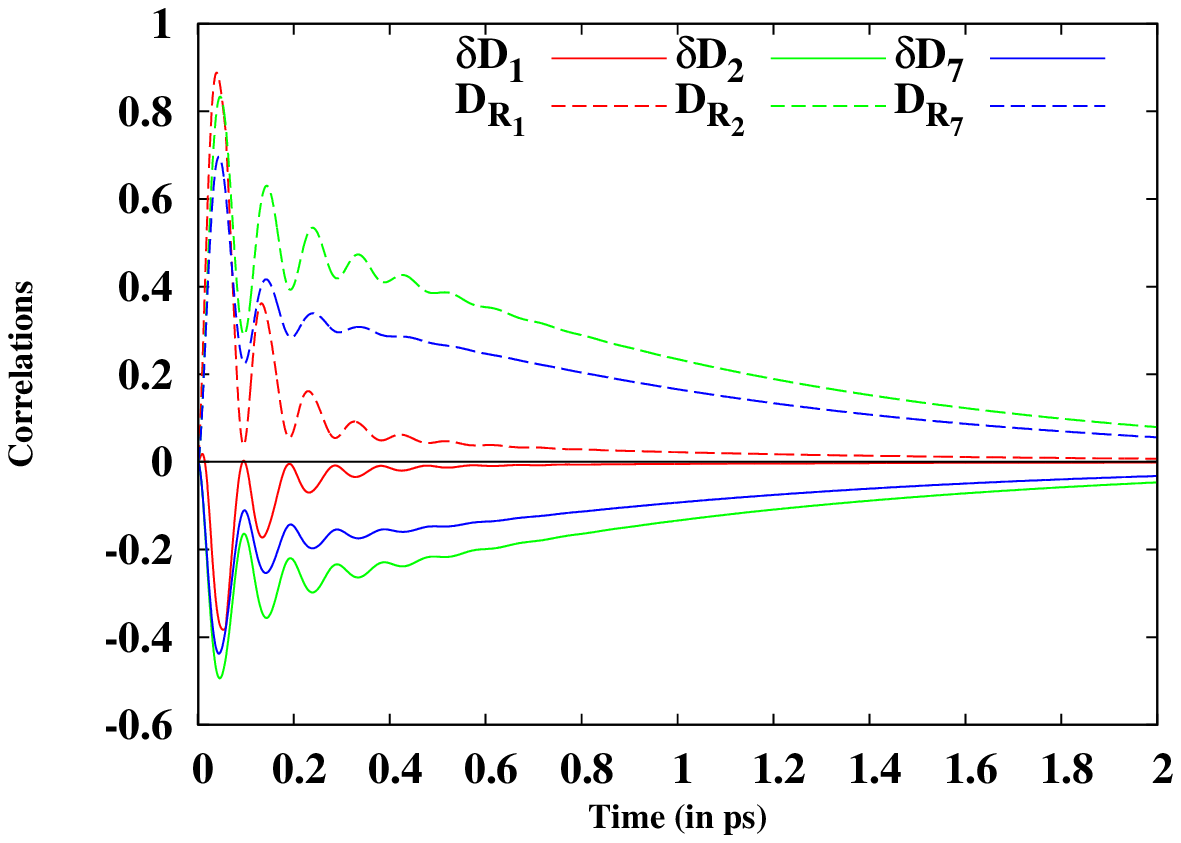}
\caption{(Color online.) Monogamy scores in the FCN in the case of dephasing mismatch. See Sec. \ref{subsec:dephmis} for the relevant parameter values. The rest of the  descriptions remain the same as in Fig. \ref{fig:msfcn_no}.}
\label{fig:dephmis}
\end{figure*}

\section{Monogamy scores and bipartition collections }
\label{sec:mngscore}
The monogamy score is a multiparty quantum correlation measure. Corresponding to any bipartite quantum correlation measure, ${\cal Q}$, the monogamy score for ${\cal Q}$ of an $N$-party quantum state $\rho_{1,2,\ldots, N}$ is defined as \cite{wootters,our_gang}
\begin{equation}
\delta {\cal Q}_i\equiv \delta {\cal Q}_i(\rho)= {\cal Q}(\rho_{i:R})-\sum_{j=1,j\neq i}^{N} {\cal Q}(\rho_{j:i}),
\end{equation}
where we have assumed the site $i$ to act as the ``nodal'' observer. Here ${\cal Q}(\rho_{i:R})$ is the quantum correlation of the entire $N$-party state $\rho_{1,2,\ldots, N}$ in the partition $i:R$, where $R$ denotes the collection of all the parties $1,2,\ldots,N$ excluding the nodal observer $i$. Note that $R$ is a function of $i$, although that has been kept silent in the notation. Here, ${\cal Q}(\rho_{j:i})$ is the quantum correlation, ${\cal Q}$, of the state $\rho_{j:i}$ obtained after tracing out all the parties of $\rho_{1,2,\ldots, N}$ except $i$ and $j$. If $ \delta {\cal Q}_{i}(\rho)$ is positive for all states  $\rho$ for all $N$, then the quantum correlation measure, ${\cal Q}$, is said to be monogamous. It has been argued that the monogamy score can act as a measure of multiparty quantum correlation \cite{wootters,our_gang}, obtained by subtracting the ``bipartite contribution'', ${\cal Q}_{R_{i}}=\sum_{j\neq i}{\cal Q}(\rho_{j:i})$, from the total quantum correlation, ${\cal Q}_{i:R} ={\cal Q}(\rho_{i:R})$ in the partition $i:R$. In this paper, we will be using monogamy scores for negativity and quantum discord, and, we will refer to them as ``negativity monogamy score'' and ``discord monogamy score'', respectively. First of all, the multiparty measures are then computable either analytically or via efficient numerical procedures. Moreover, negativity and quantum discord fall on different sides of the broad division in the space of bipartite quantum correlation measures into  entanglement-separability measures and information-theoretic ones. The multiparty quantum correlations generated thereof will therefore lead to the identification and understanding of a breadth of features on sharability of quantum correlations in the system considered. 

Along with using monogamy scores as multiparty quantum correlation measures, we also use collections of quantum correlations in different bipartitions for the same purpose. Given an $N$-party quantum state $\rho_{1,2,\ldots,N}$, we consider a bipartite quantum correlation ${\cal Q}$, as for the monogamy scores. We then consider the collection ${\cal Q}_{i:R}$ for all $i$. The collection provides information about how the quantum correlation, ${\cal Q}$, is shared between different single party and the rest bipartitions, and hence gives us an understanding of the multiparty quantum correlation in the entire state. This is somewhat similar to the concept of area law (or its violation) where entanglement of different bipartitions of the entire system is utilized to gather information about  the character of a many-body system, including quantum phase transitions \cite{area_law}.

\section{Dynamics of bipartite and multipartite quantum correlations}
\label{sec:mngfmofn}

In this section, we will first present our results on bipartite and multiparty quantum correlations in the fully connected network. Then we apply the concept of monogamy of quantum correlations to understand the exciton transport in FMO complexes. 

\subsection{Fully connected network}
\label{sec:mngfcn}

We consider a $7$-site FCN model for the discussion. In a FCN, all the hopping amplitudes, denoted by $v_{ij}$, are taken to be equal \cite{caruso_2009, ekert}. The evolution of the density matrix under dissipation, dephasing, and the sink operator is obtained by using Eq. (\ref{eq:evolvedm}). We choose site $7$ as the preferred one, which is connected to the sink at site $8$. The evolution starts off from an initial state, which in this case is taken to be $\arrowvert 1 \rangle$.

\subsubsection{Clean case}
\label{subsec:nomis}

Consider the Hamiltonian given in Eq. (\ref{eq:NM_hamil}) and further assume that  $\Gamma_j$ and $\gamma_{j}$ are zero for all $j=1,\ldots,N$.  Physically it means that the system is dissipation- and dephasing-free, and excitons created in the system are not destroyed due to any environmental effect. Rather, they only decay to the sink (site $8$) from site $7$. Such FCN models will be referred to as ``clean'' cases throughout the manuscript.  For simplicity, all $\omega_j$ are chosen to be zero. The rate of transfer of energy to the sink is $\Gamma_{N+1}$ which is chosen to be $50 \ \text{cm}^{-1}$. We also choose $\hbar v_{ij} = J = 50 \ \text{cm}^{-1}$. Starting with the initial state where only one electronic exciton is present at the first site, creation and annihilation of excitons take place as the excitation moves from one site to other sites over a time period which is of the order of a picosecond, and finally the energy is transported from the first site to the sink, numbered $8$, via the dynamics of the exciton.

To consider the monogamy scores, we have to identify a nodal observer. The different situations correspond to cases  when the nodal observers are located at sites $1$, $2$ or $7$. (Note here that the sites $2 - 6$ are equivalent in this case.) We begin with the case when the nodal observer is at site $1$. In this case, the negativity monogamy score is given by
\begin{equation}
\delta{ N}_1={ N}_{1:R}-\sum_{j=2}^{7}{ N}_{j:1}.
\end{equation} 
\begin{figure*}
\includegraphics[width=0.3\linewidth]{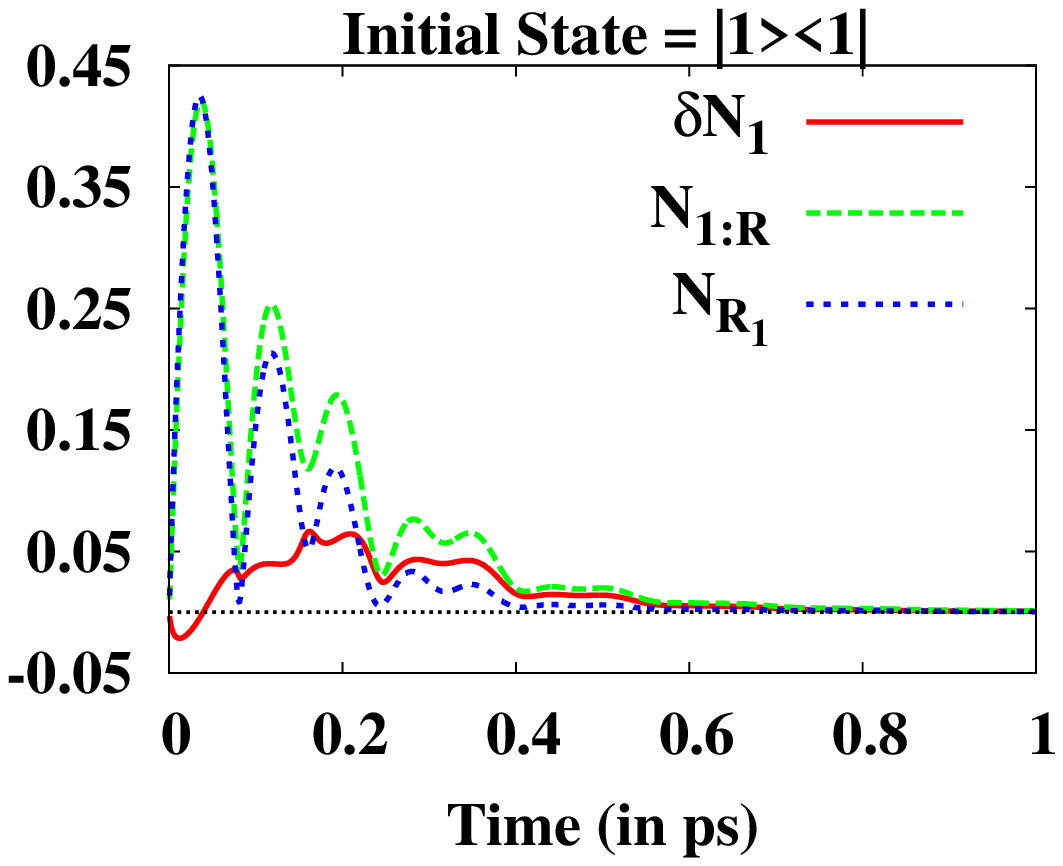}
\includegraphics[width=0.3\linewidth]{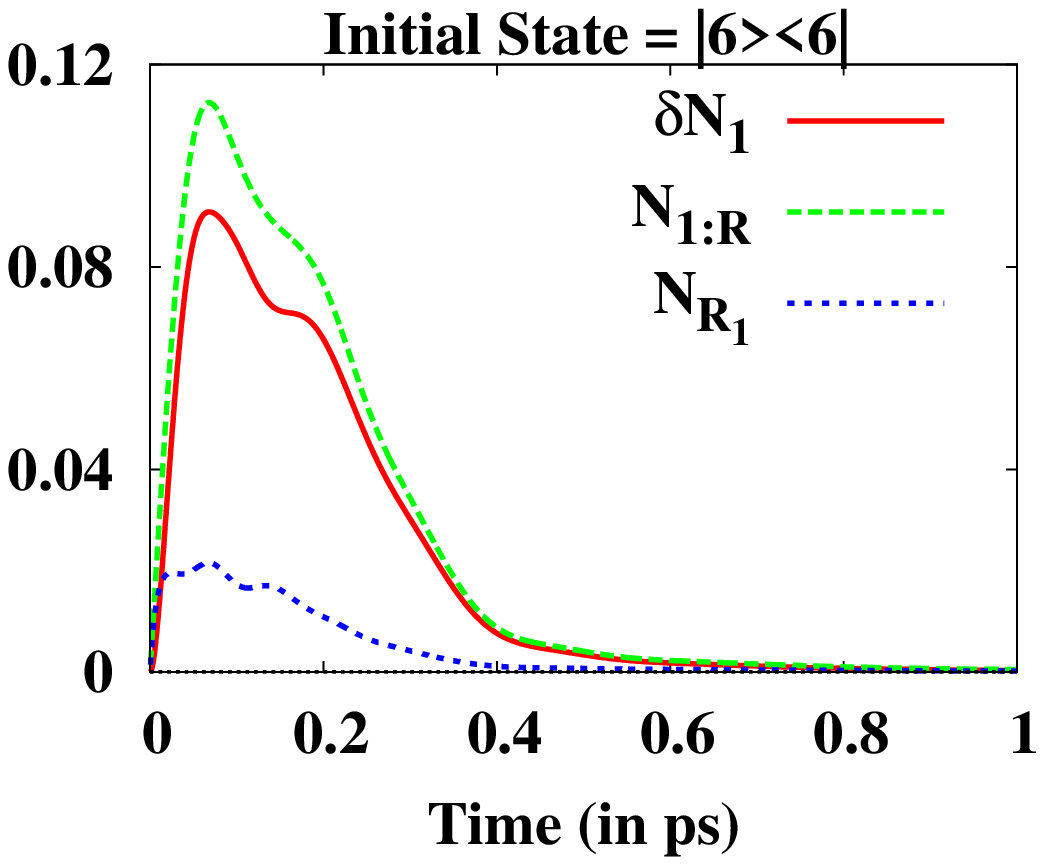}
\includegraphics[width=0.3\linewidth]{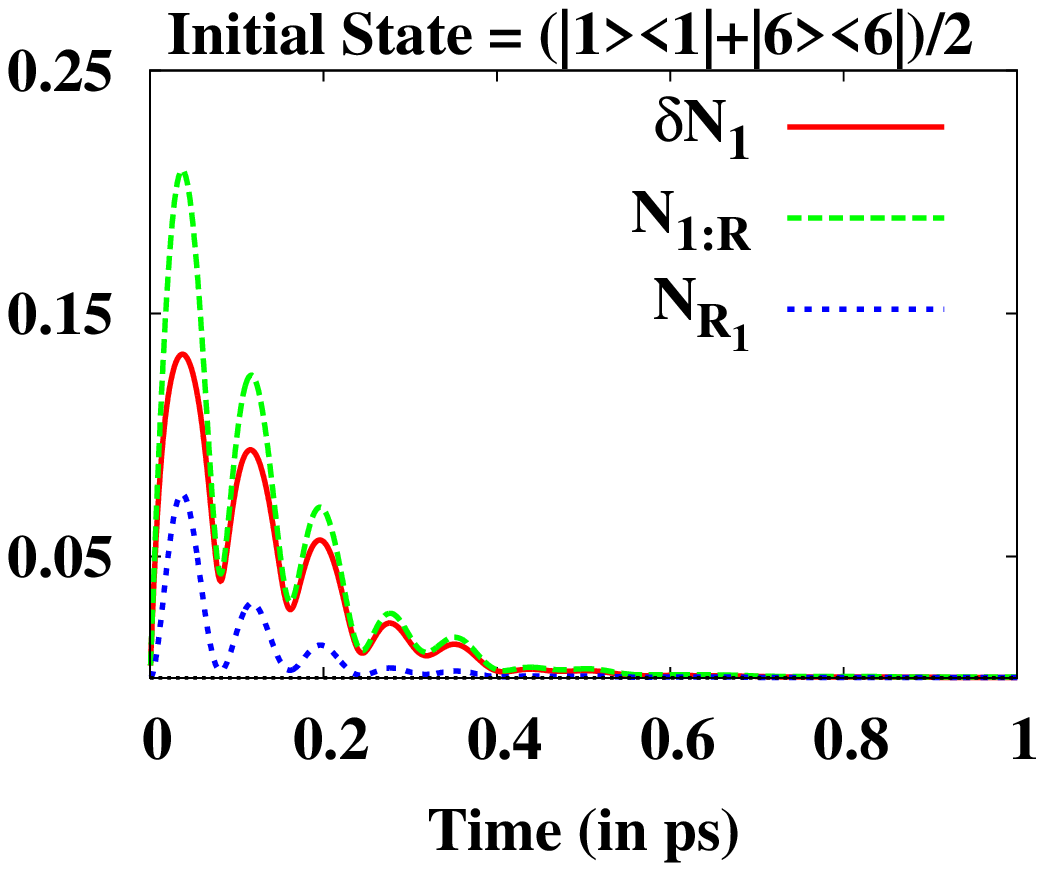}
\\
\includegraphics[width=0.3\linewidth]{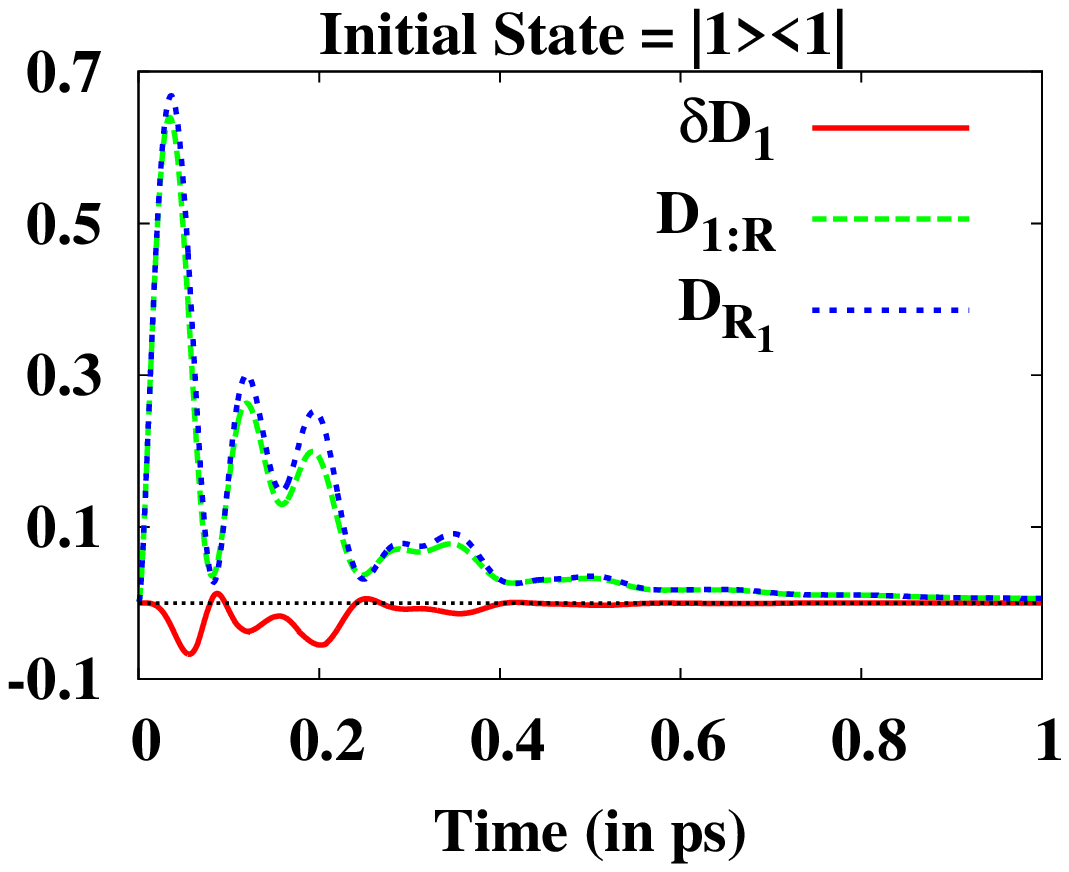}
\includegraphics[width=0.3\linewidth]{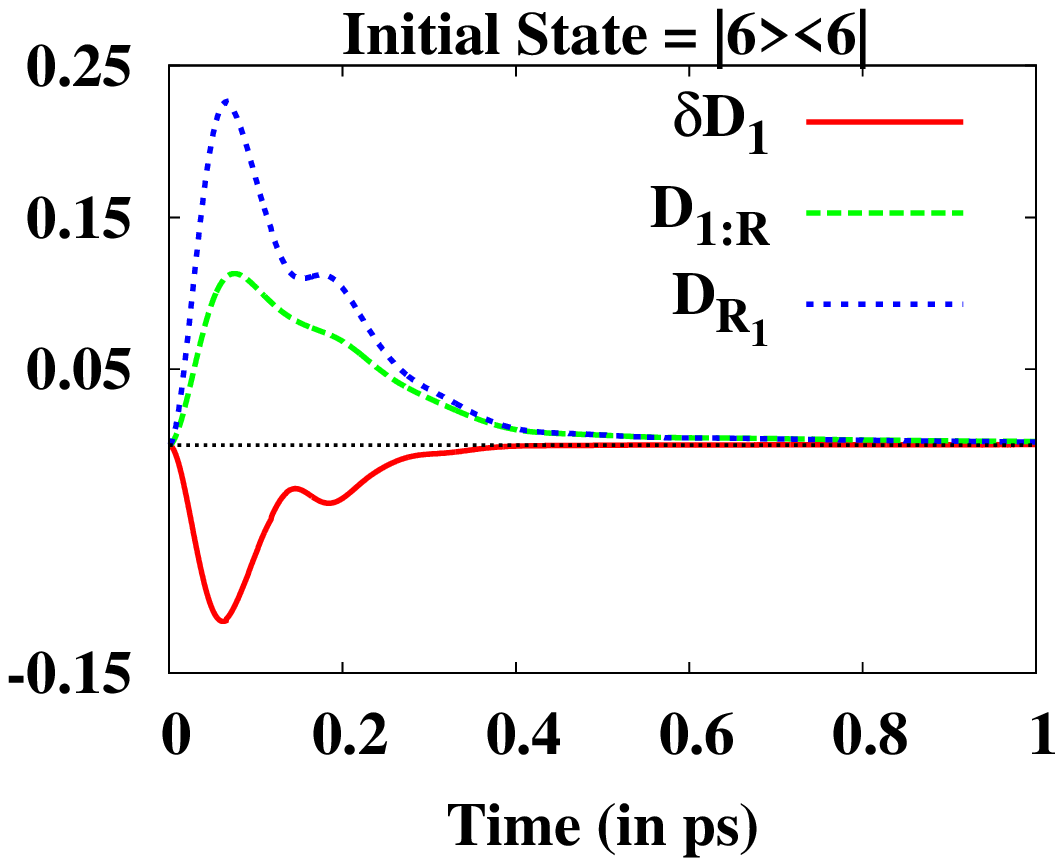}
\includegraphics[width=0.3\linewidth]{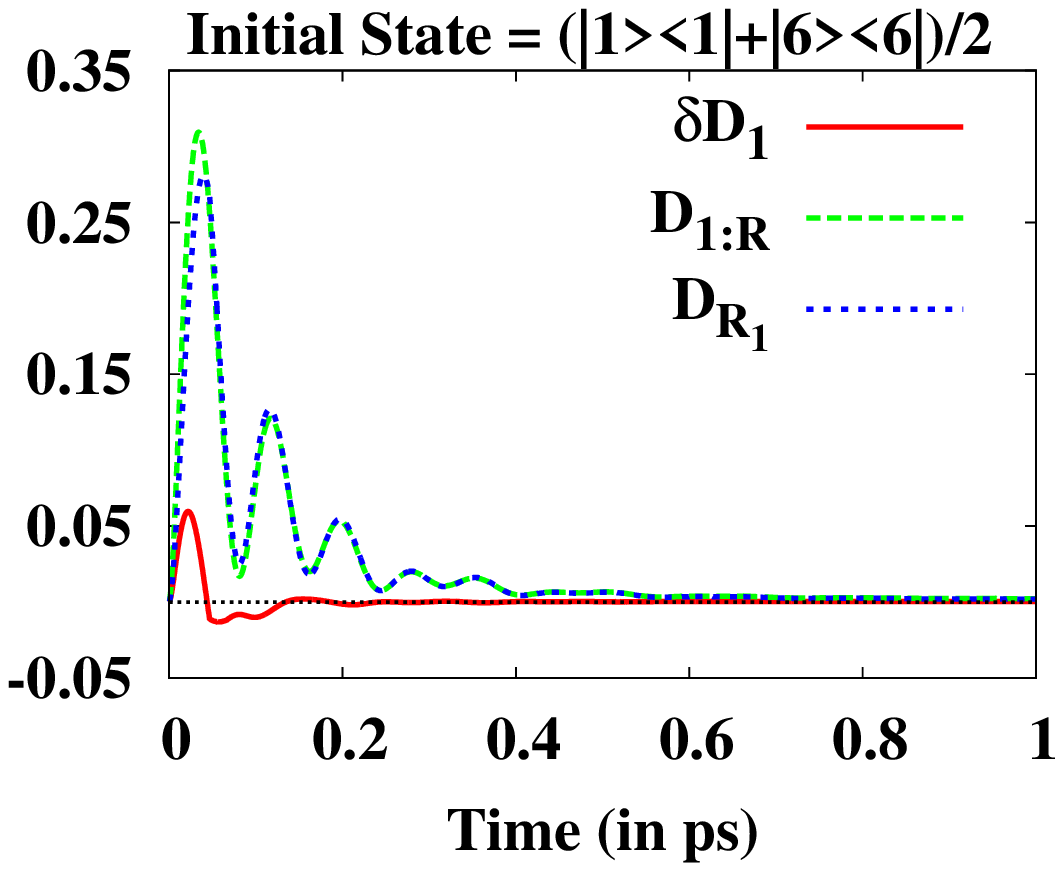}
\caption{(Color online.) Monogamy scores in the FMO complex with site $1$ as the nodal observer. The top row of panels is for negativity as the quantum correlation measure, while the bottom row is for quantum discord. The three columns of panels are (from left to right) for the initial state as $\arrowvert 1\rangle\langle 1 \arrowvert$, $\arrowvert 6\rangle\langle 6 \arrowvert$, and their equal mixture respectively. The line of zero ordinate, wherever present is for pointing out the distinction between negative and positive values of the negativity and discord monogamy scores with time. The qualitative behavior is very similar when the site $2$ is used as the nodal observer. The ordinates in the top panels are in ebits, while those in the bottom one are in bits.}
\label{fig:fmo1}
\end{figure*}

\noindent A  similar monogamy score can be defined for quantum discord ($\delta D_1$), and for other sites as nodal observers. The monogamy score for negativity in this case is depicted in Fig. \ref{fig:msfcn_no}. The evolution of the system is monitored for a period of $10$ ps, but exhibited in the figure only up to $2$ ps. After $2$ ps, the oscillations in the correlations have either vanished or are steadily decreasing. The monogamy score, $\delta N_1$, remains negative throughout the evolution. A non-zero value of the monogamy score indicates the multipartite nature of entanglement in the state \cite{wootters}. Similar features are also observed for the discord monogamy score when site $1$ is the nodal observer. It is to be noted here that while $\delta  N_1$ saturates after $1.2$ ps, $\delta D_1$ exhibits significant oscillations even after $2$ ps. The features for negativity and discord monogamy scores are broadly similar in the case when the site $2$ is considered to be the nodal observer. A notable difference is that $\delta N_2 >0$ for all times, so that the dynamically evolved states are monogamous for all times.
See Fig. \ref{fig:msfcn_no}.  

The story of entanglement sharability among the sites of the FCN is not the same when the site $7$ is considered as the nodal observer. The evolved state, in this case, exhibits both monogamous and non-monogamous behavior with time with respect to entanglement. See Fig. \ref{fig:msfcn_no}.
Initially, the negativity monogamy score oscillates between positive and negative values and after some picoseconds, the monogamy score vanishes. On the other hand the discord monogamy score with respect to site $7$ as the nodal observer is always negative. Interestingly therefore, the discord monogamy scores of the evolved state is negative irrespective of the site chosen as the nodal observer.

The population transfer for the clean case can be calculated explicitly, and for large time, i.e. for $t\rightarrow \infty$, it is given by 
\begin{equation}
p_{sink}(\infty)=\frac{1}{N-1}.
\end{equation}
For the $7$-site network model, $p_{sink}\approx 0.1667$. So, the rest of the population remains in the network sites and contributes to the quantum correlations and monogamy scores, except for site $7$, for which  the bipartite quantum correlations, $N_{R_7}$ and $D_{R_7}$, as well as the monogamy scores, $\delta N_7$ and $\delta D_7$, decay to zero after sufficient time. This is because the population at the site (site $7$) connected to sink is zero after sufficient time. See Fig. \ref{pop_FCN}.

\subsubsection{Energy mismatch}
\label{subsec:emis}
In this subsection, we will consider the case when the on-site energy of one of the sites (in our case, site $1$) is different from the other sites in the FCN. We have chosen $\hbar \omega_1 = 50 \ \text{cm}^{-1}$. It has been argued that the energy mismatch at one or more sites  introduces a ``static disorder'' in the FCN model and is responsible for an increase in transportation efficiencies. We have seen that having one or more sites with different energies, the transportation efficiency of the FCN channel can be made to unity, even without any dephasing. We find that the behavior of sharability of quantum correlations changes drastically due to such energy mismatch. In particular, the negativity monogamy score, $\delta N_1$, now oscillates between being monogamous and non-monogamous for $0\leq t\leq 1.4$ (in ps), and vanishes thereafter. See Fig. \ref{fig:msfcn_en}. In the clean case, $\delta N_1$ remains non-monogamous for all time. Another important change in behavior is that the quantum correlations $\mathcal{Q}_{R_i}$ and the $\delta \mathcal{Q}_i$ for all $i$ decay to zero with time. This is in contrast to the behavior of these quantities in the clean case, where many of them had converged to non-zero steady values, though $\mathcal{Q}_{R_7}$ and $\delta \mathcal{Q}_7$ last longer in the clean case. Just like for the clean case, the discord monogamy scores ($\delta D_i$) are non-monogamous throughout the dynamics.

\begin{figure*}
\includegraphics[width=0.3\linewidth]{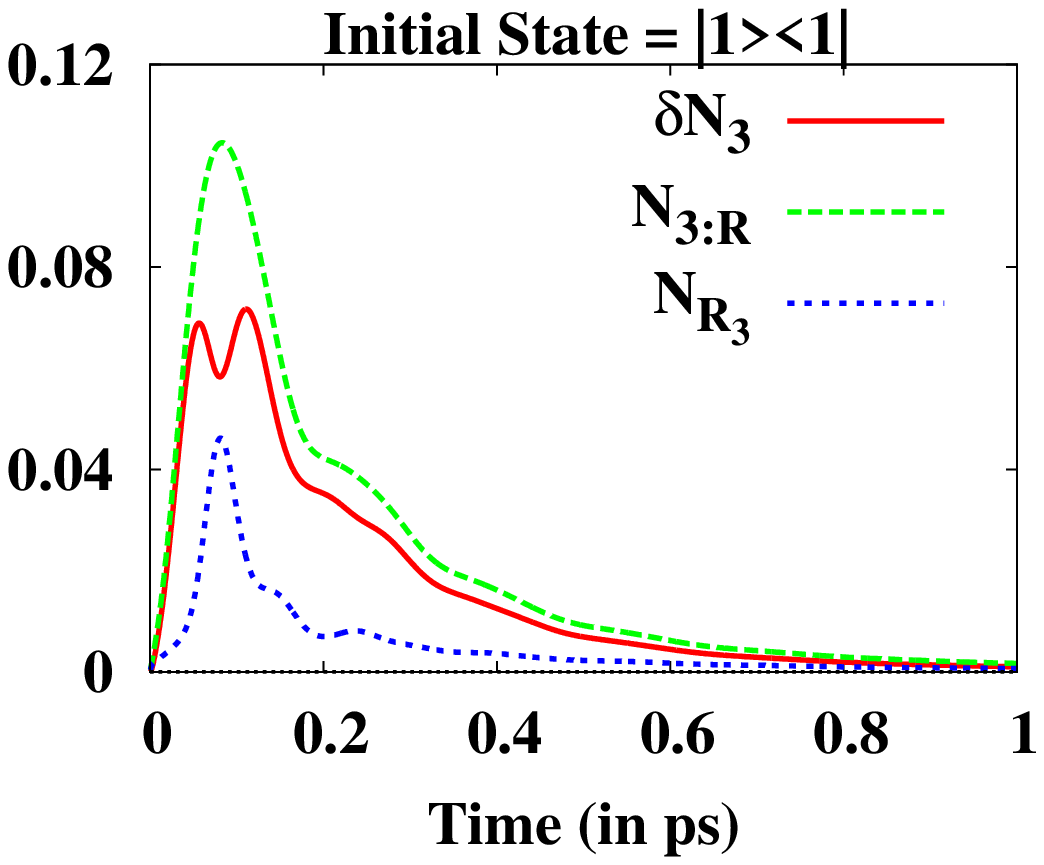}
\includegraphics[width=0.3\linewidth]{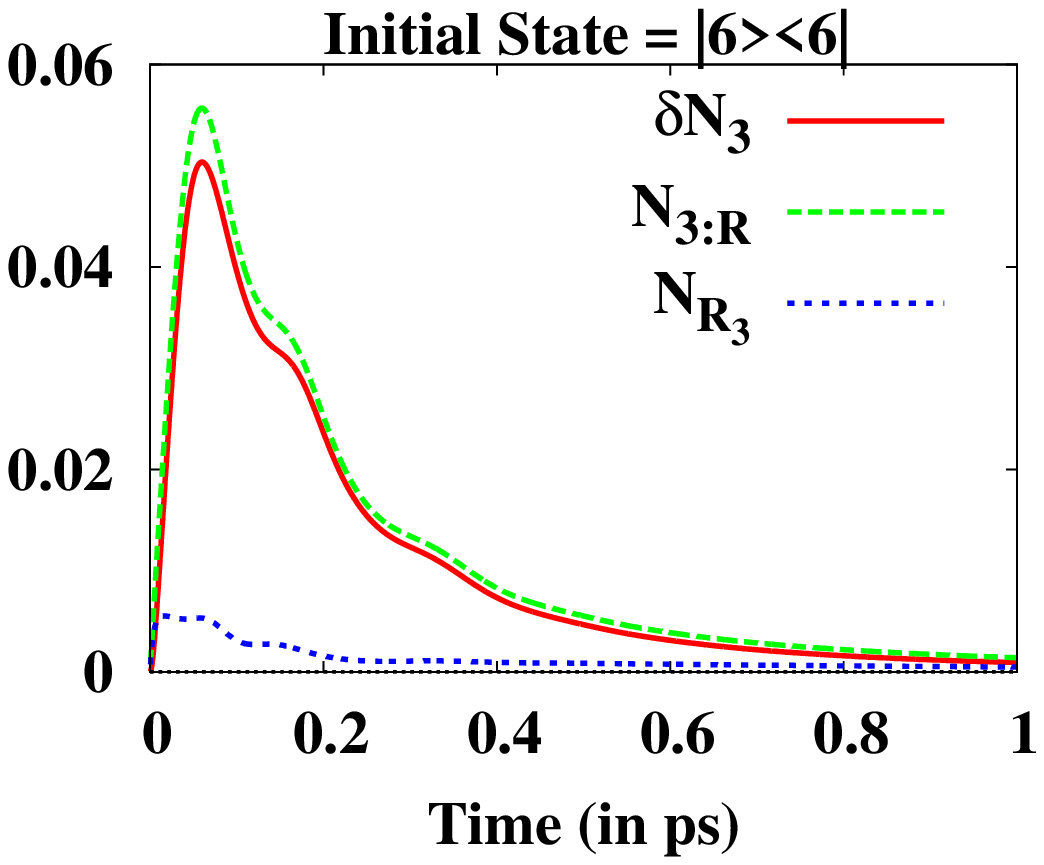}
\includegraphics[width=0.3\linewidth]{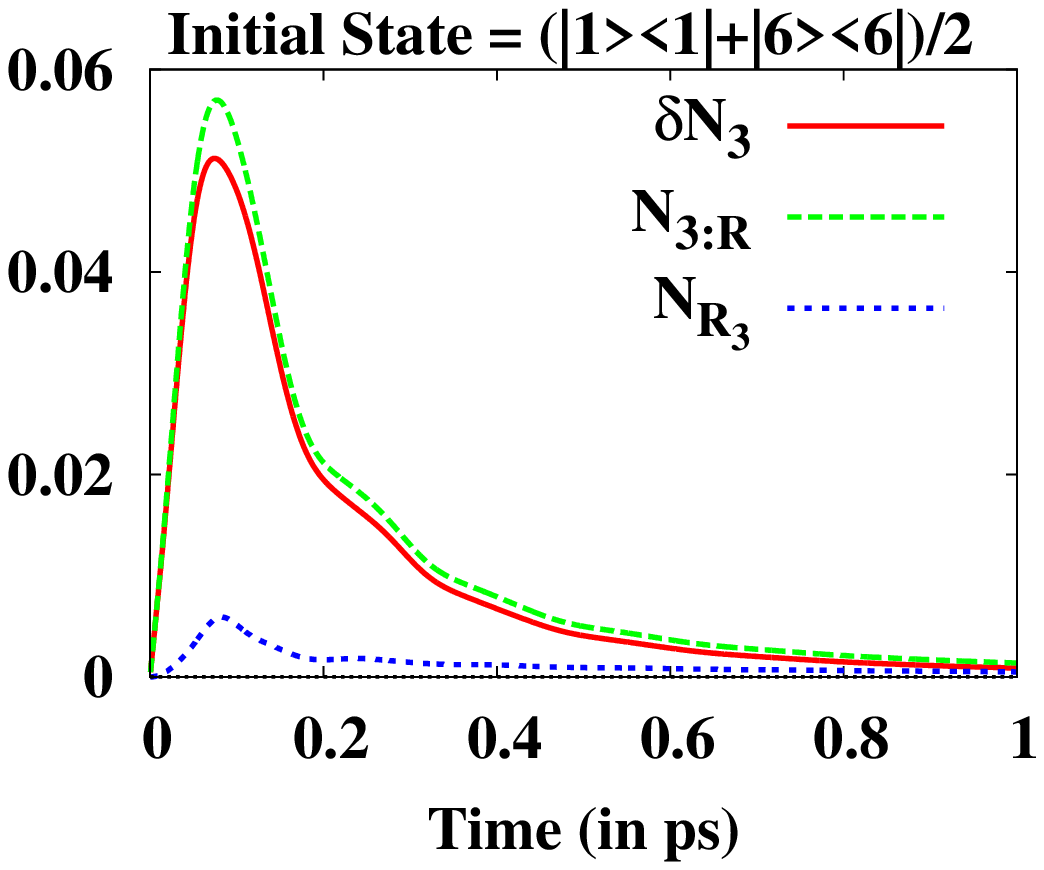}
\\
\includegraphics[width=0.3\linewidth]{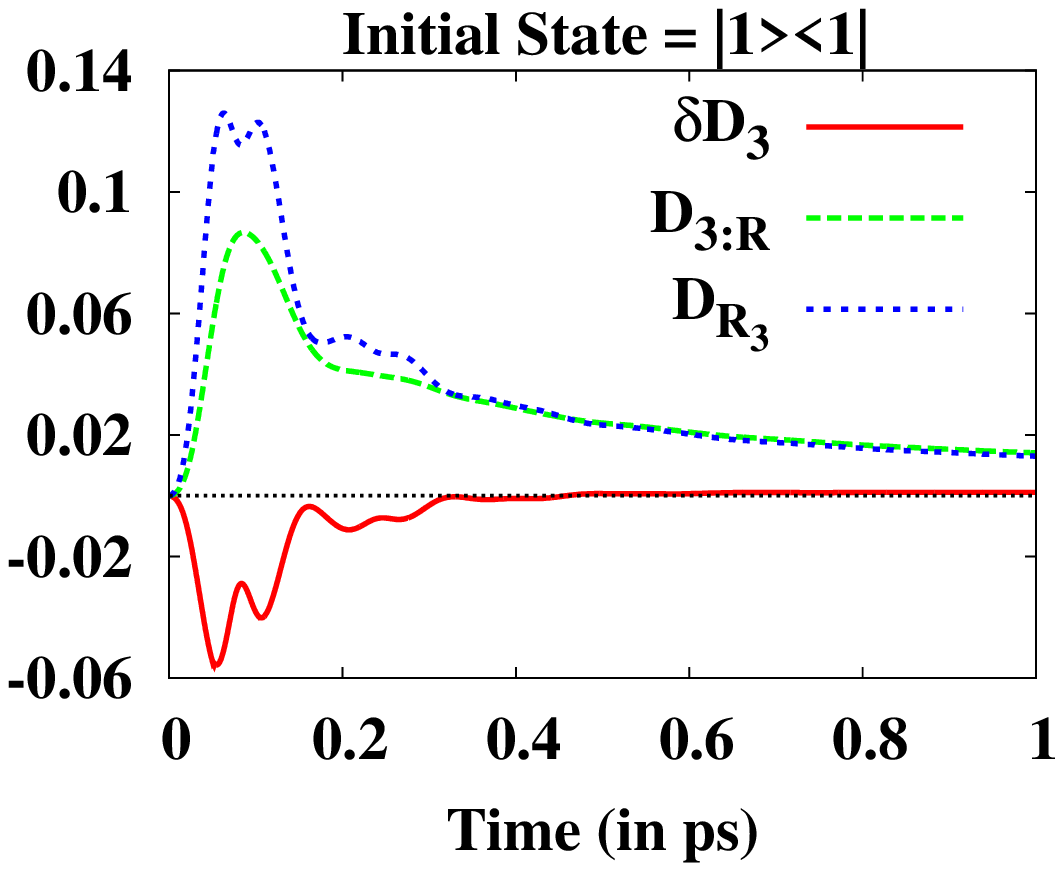}
\includegraphics[width=0.3\linewidth]{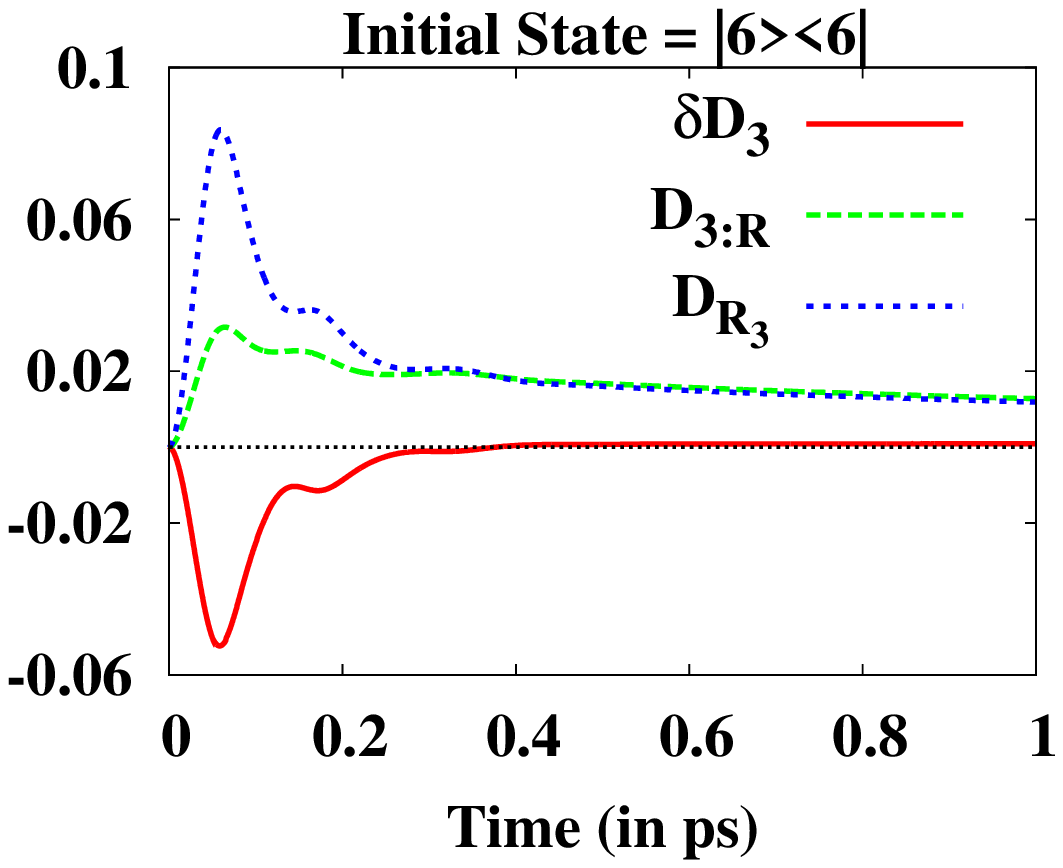}
\includegraphics[width=0.3\linewidth]{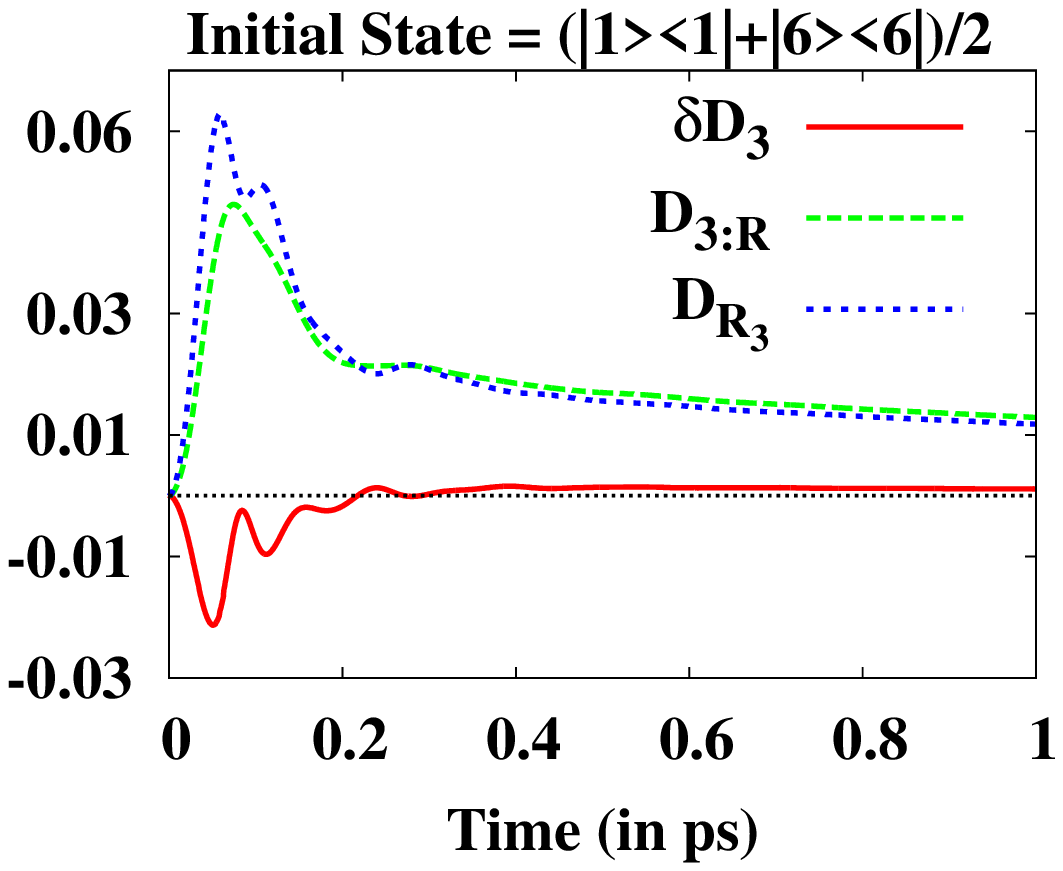}
\caption{(Color online.) Monogamy scores in the FMO complex with the site $3$ as the nodal observer. The remaining descriptions are the same as in Fig. \ref{fig:fmo1}, except that the qualitative features remain the same when the site $4$ or the site $7$ is used as the nodal observer. }
\label{fig:fmo2}
\end{figure*}

\subsubsection{Dephasing mismatch}
\label{subsec:dephmis}
Now let us consider the situation where one of the dephasing parameters is non-zero and rest are 
set to zero. We choose $\gamma_1 = 50 \ \textrm{cm}^{-1}$. The on-site energies are all set to zero. It is found that an increase in the  dephasing  increases the transfer of population in this model. The corresponding plots for quantum correlations are shown in Fig. \ref{fig:dephmis} for different sites as the nodal observers. Just like in the clean case and in the case of energy mismatch, the panels in the figure exhibits bipartite contributions as well as monogamy scores of quantum correlations with respect to the different nodal observers. The bipartite contribution is initially high in the case of site $1$ as nodal observer, shown by a dashed red line in Fig. \ref{fig:dephmis}, and the state of the whole system remains non-monogamous for around $0.05$ ps, and after that it becomes monogamous. The negativity monogamy scores are  positive for other sites as nodal observers for almost all time except a short period at the beginning. The discord monogamy scores on the other hand remain negative for all the sites as nodal observers (see Fig. \ref{fig:dephmis}). Unlike the clean case, the discord monogamy score with site $1$ as the nodal observer decays faster than the one with site $7$ as the nodal observer. In both the cases, if the site $2$ is taken as the nodal observer quantum correlation sustain for an even longer period. We will see that such analysis of quantum correlations is useful to detect energy transfer pathways in network models (especially in the FMO complex).

\begin{figure*}[t]
\includegraphics[width=0.3\linewidth]{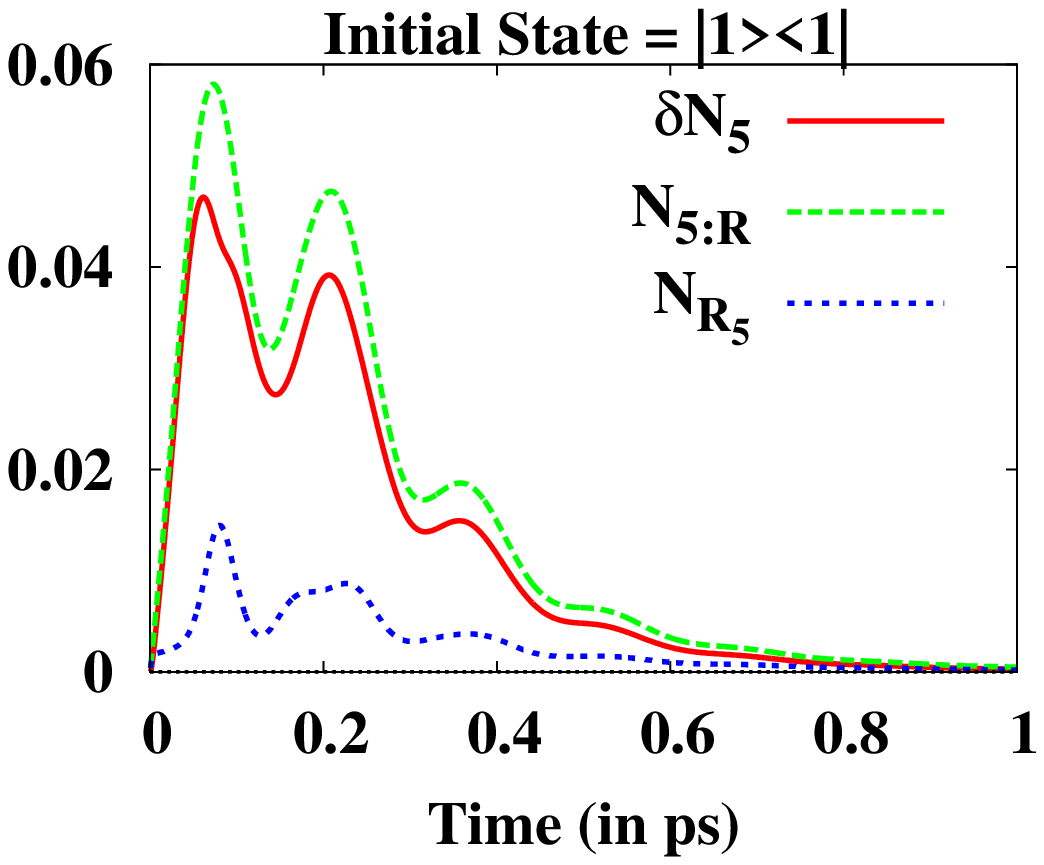}
\includegraphics[width=0.3\linewidth]{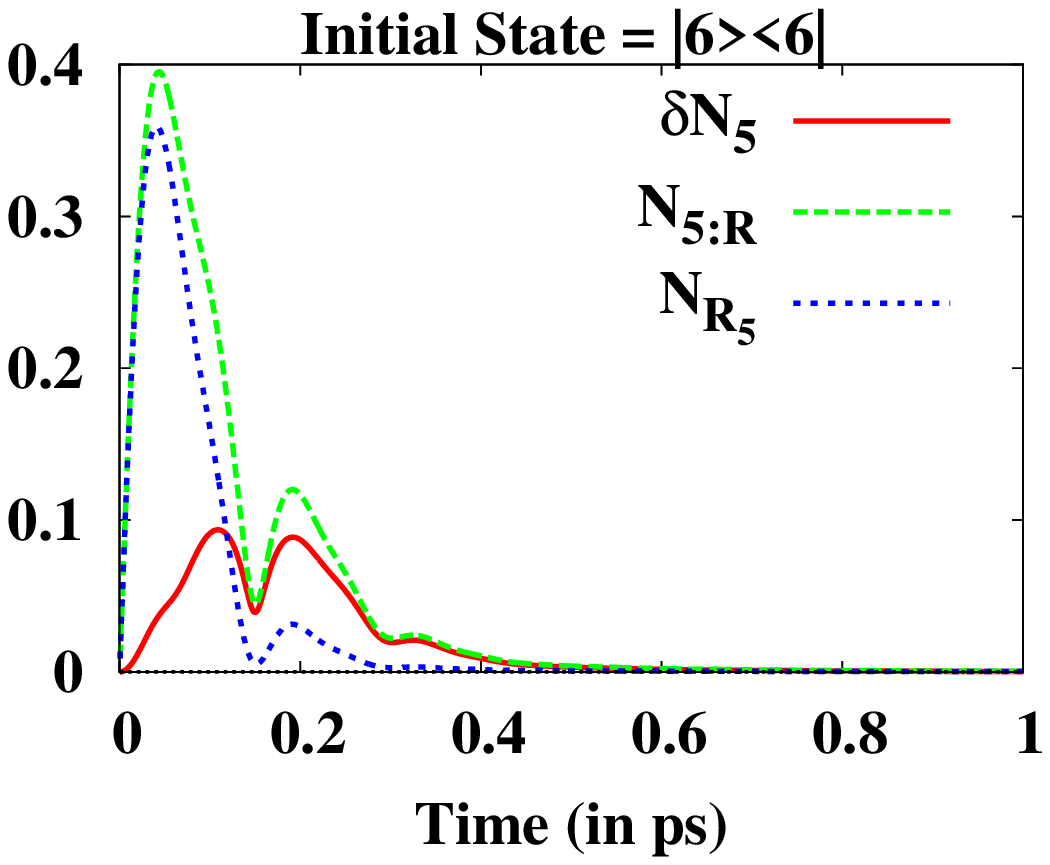}
\includegraphics[width=0.3\linewidth]{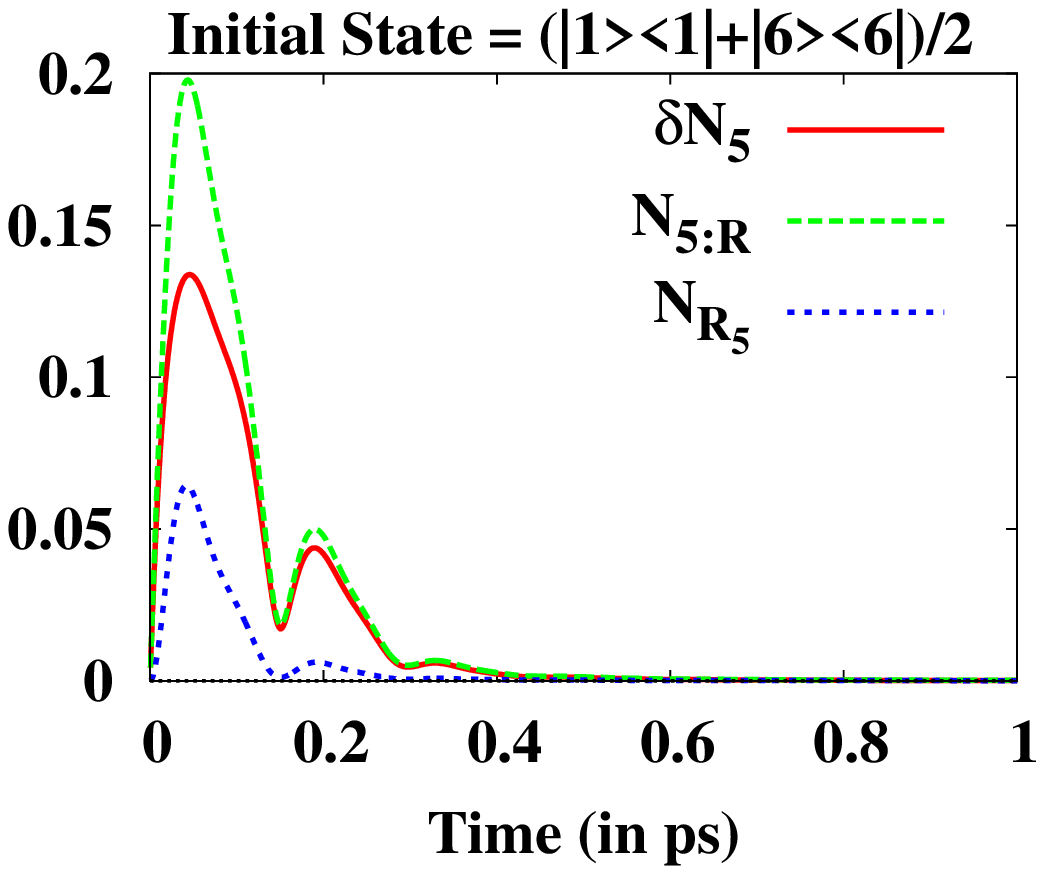}
\\
\includegraphics[width=0.3\linewidth]{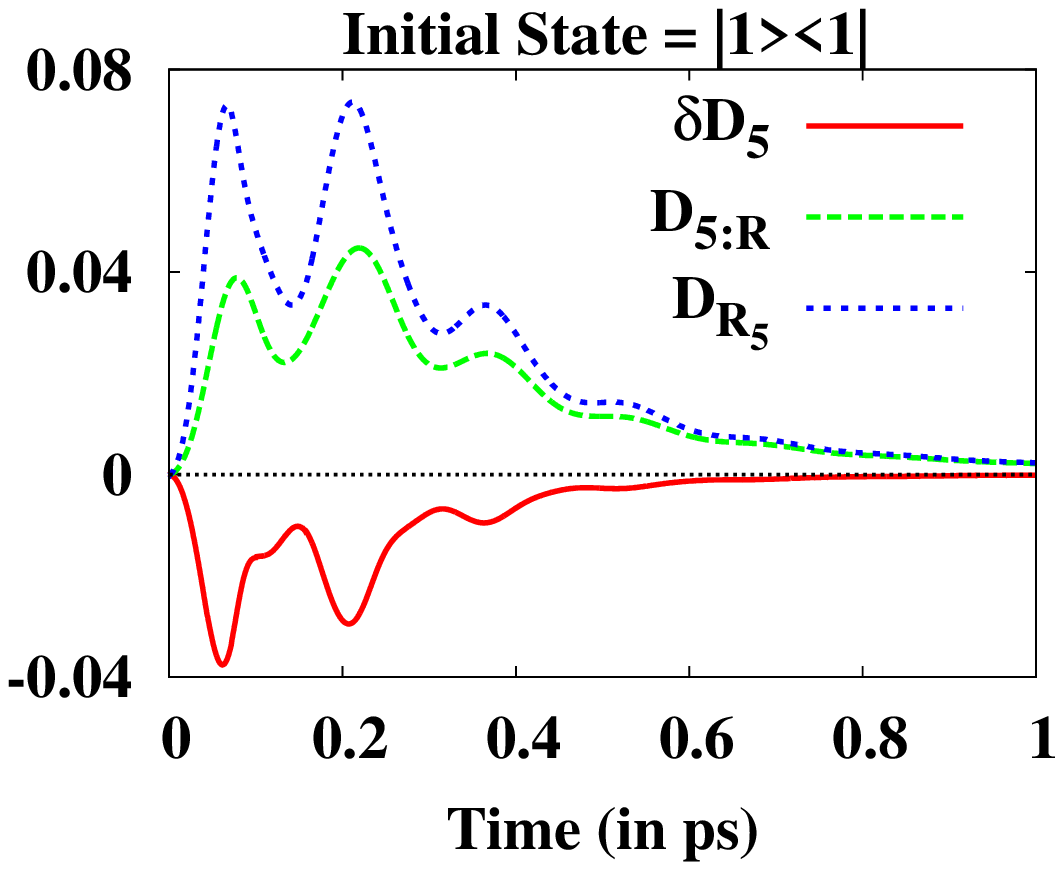}
\includegraphics[width=0.3\linewidth]{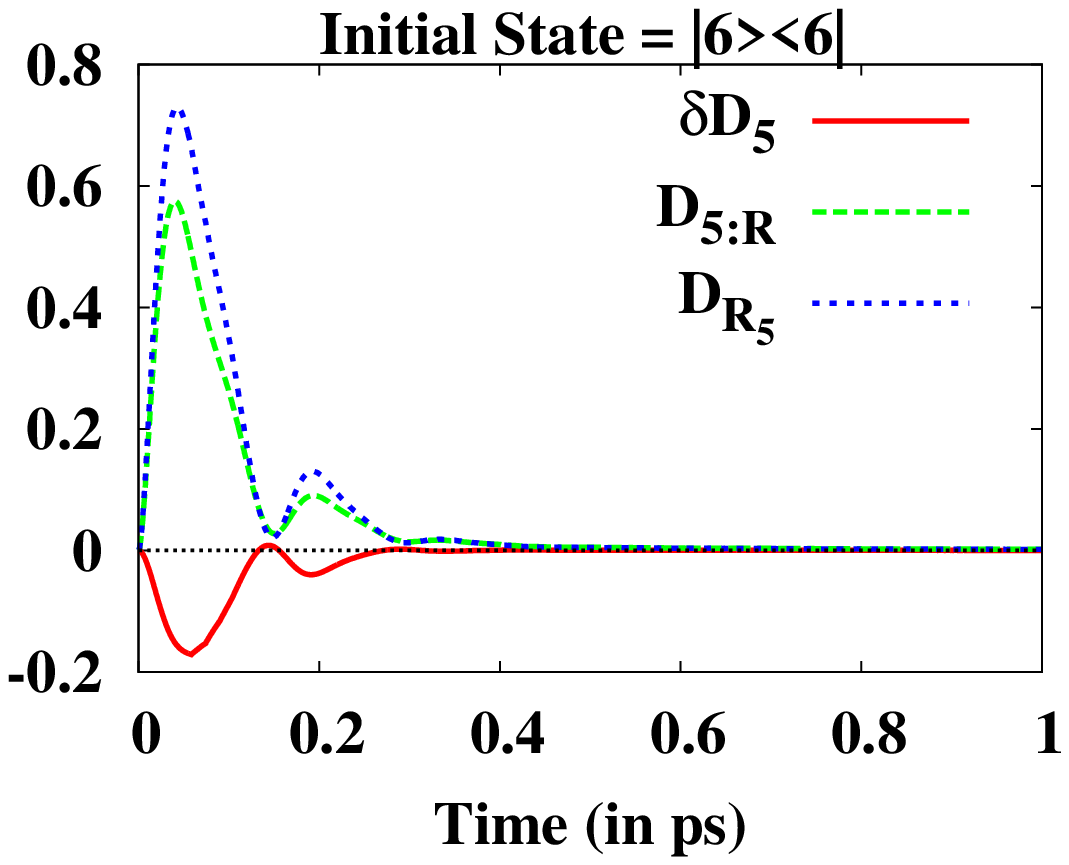}
\includegraphics[width=0.3\linewidth]{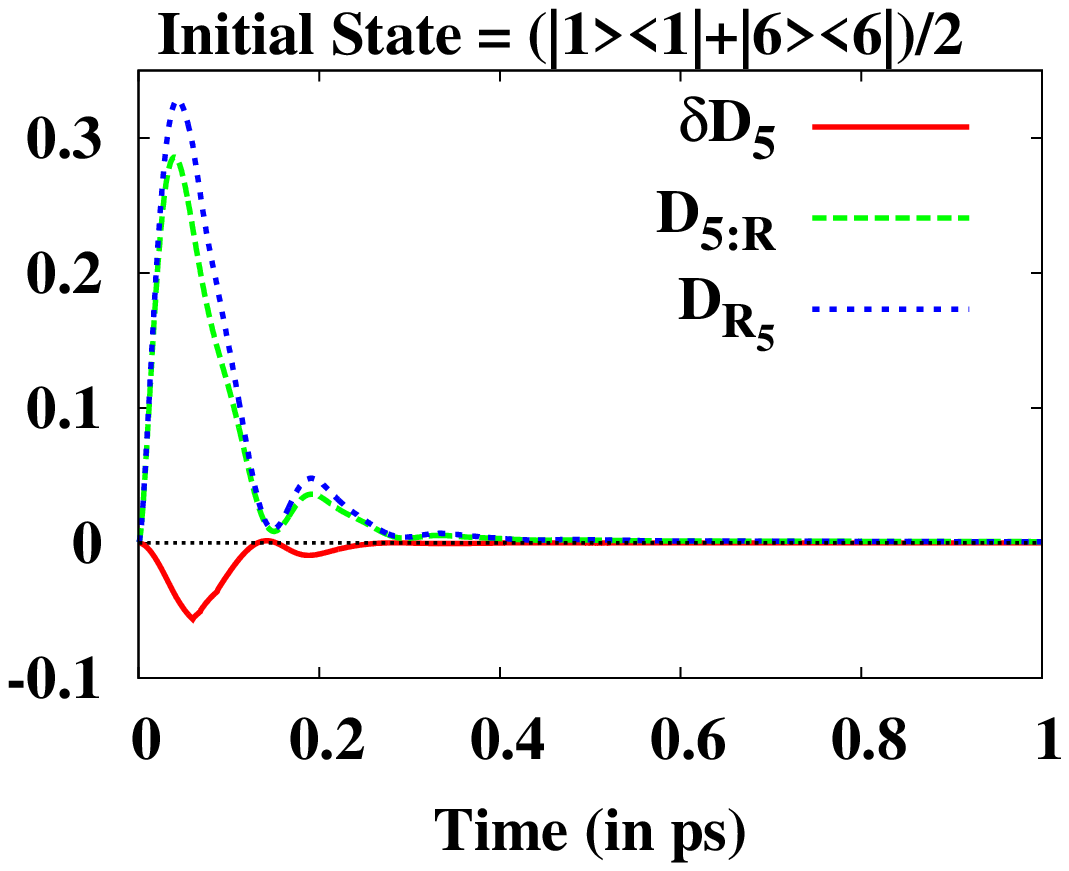}

\includegraphics[width=0.3\linewidth]{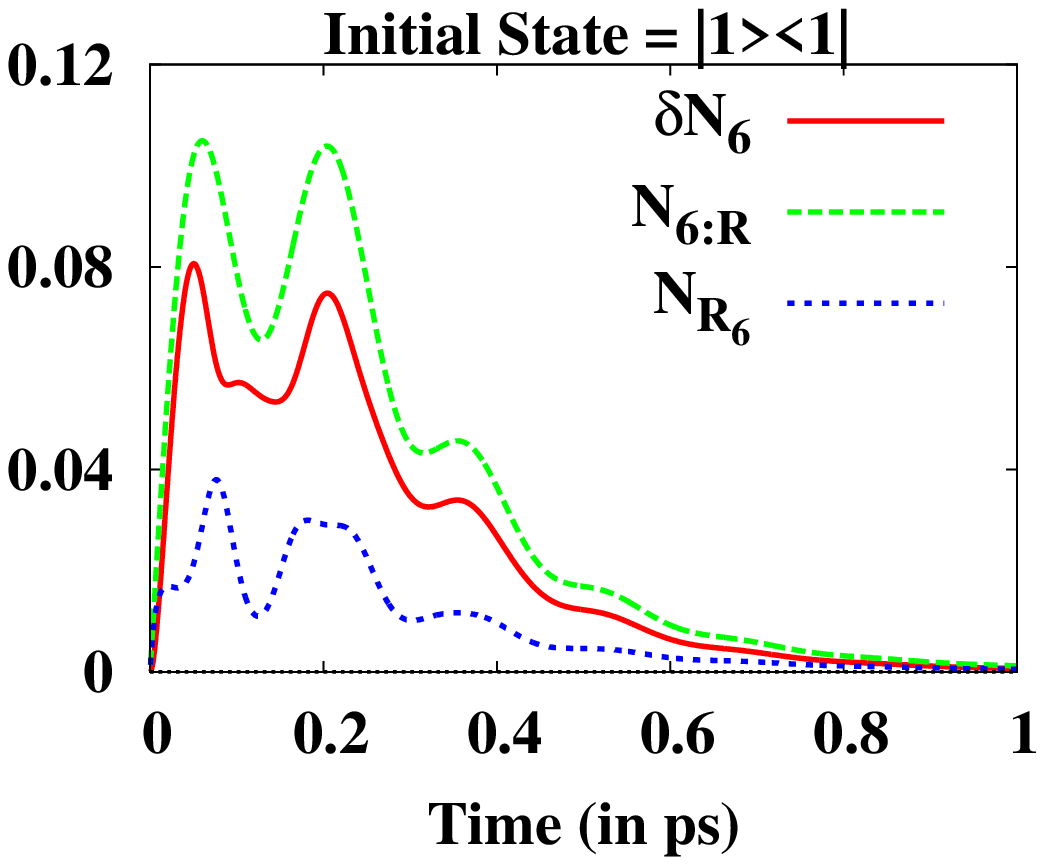}
\includegraphics[width=0.3\linewidth]{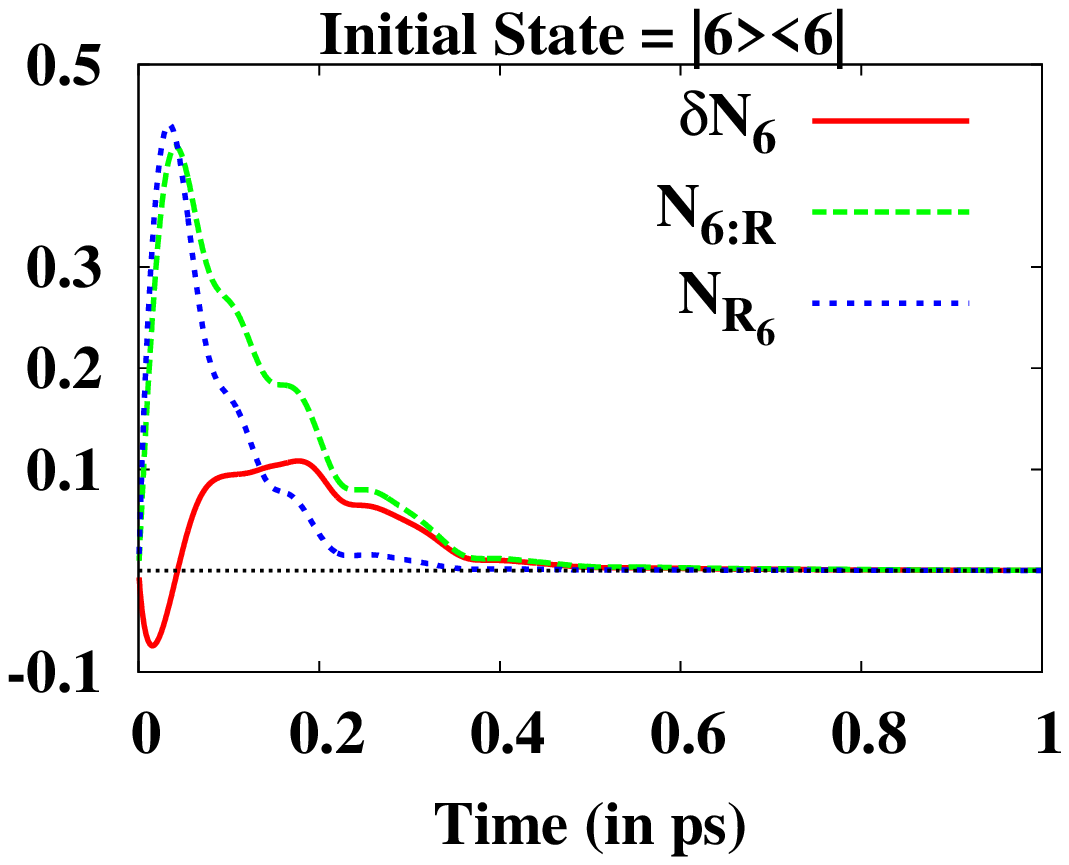}
\includegraphics[width=0.3\linewidth]{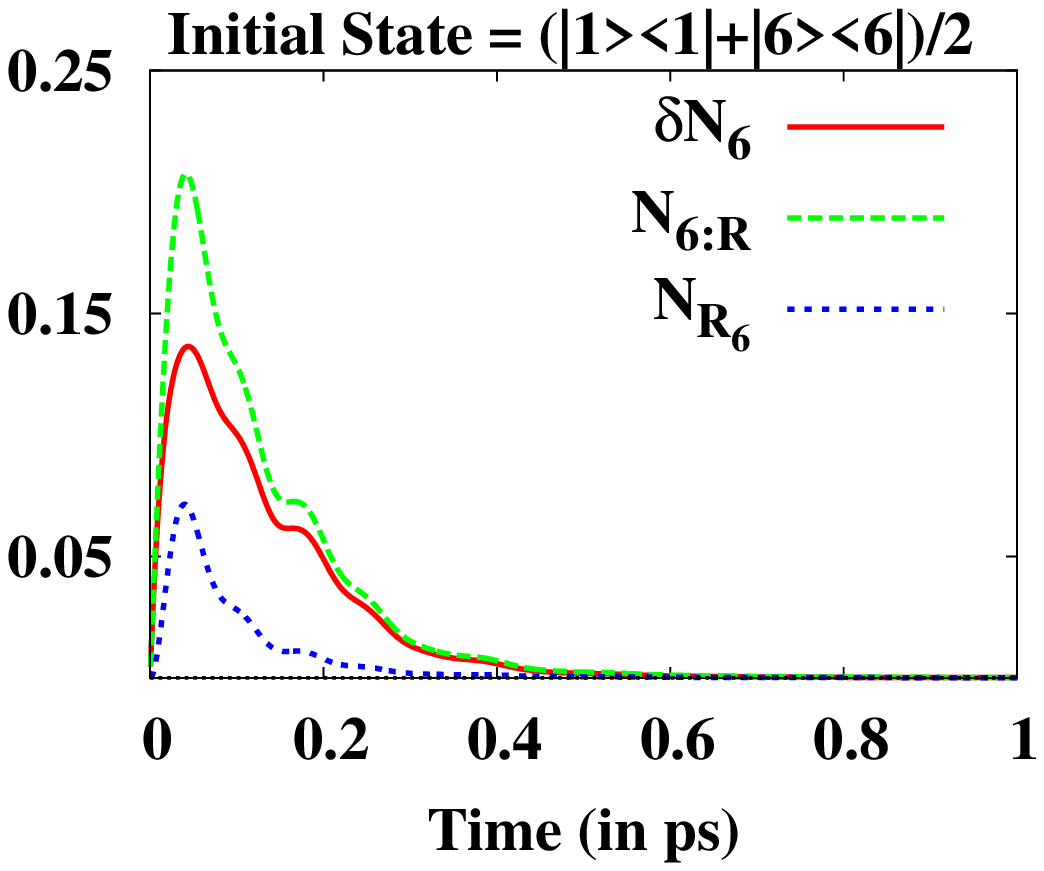}
\\
\includegraphics[width=0.3\linewidth]{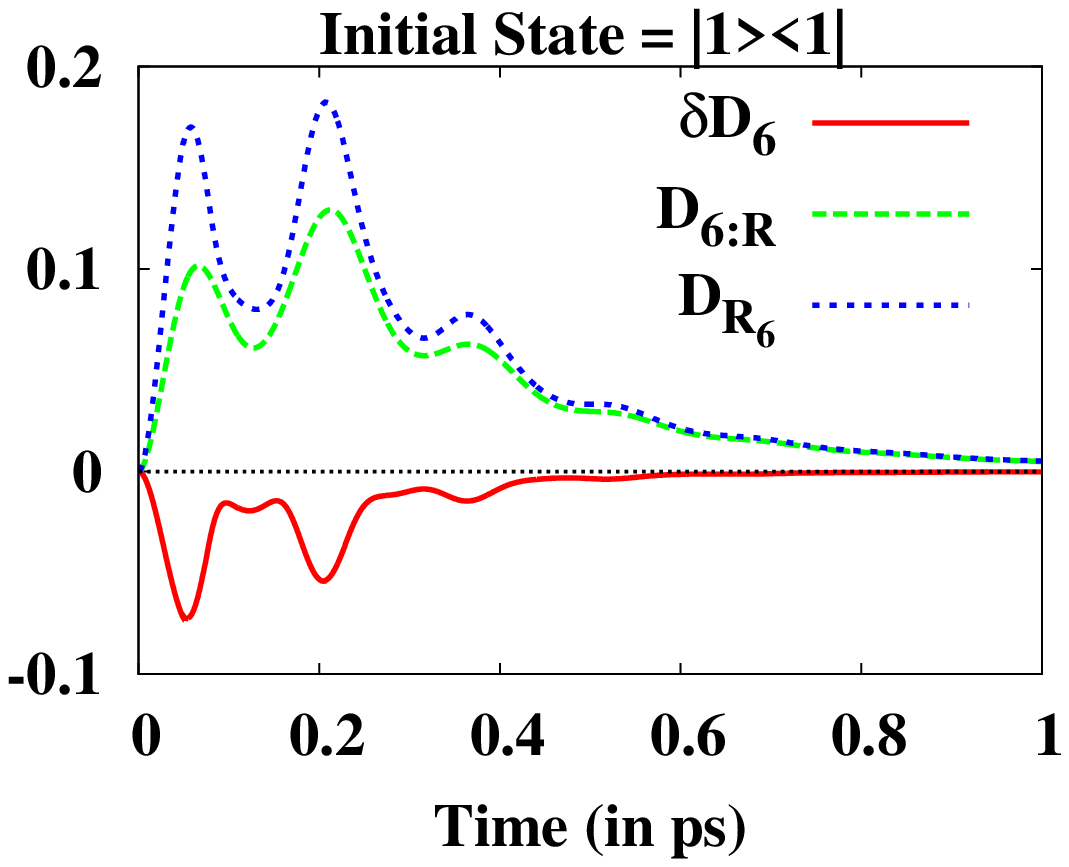}
\includegraphics[width=0.3\linewidth]{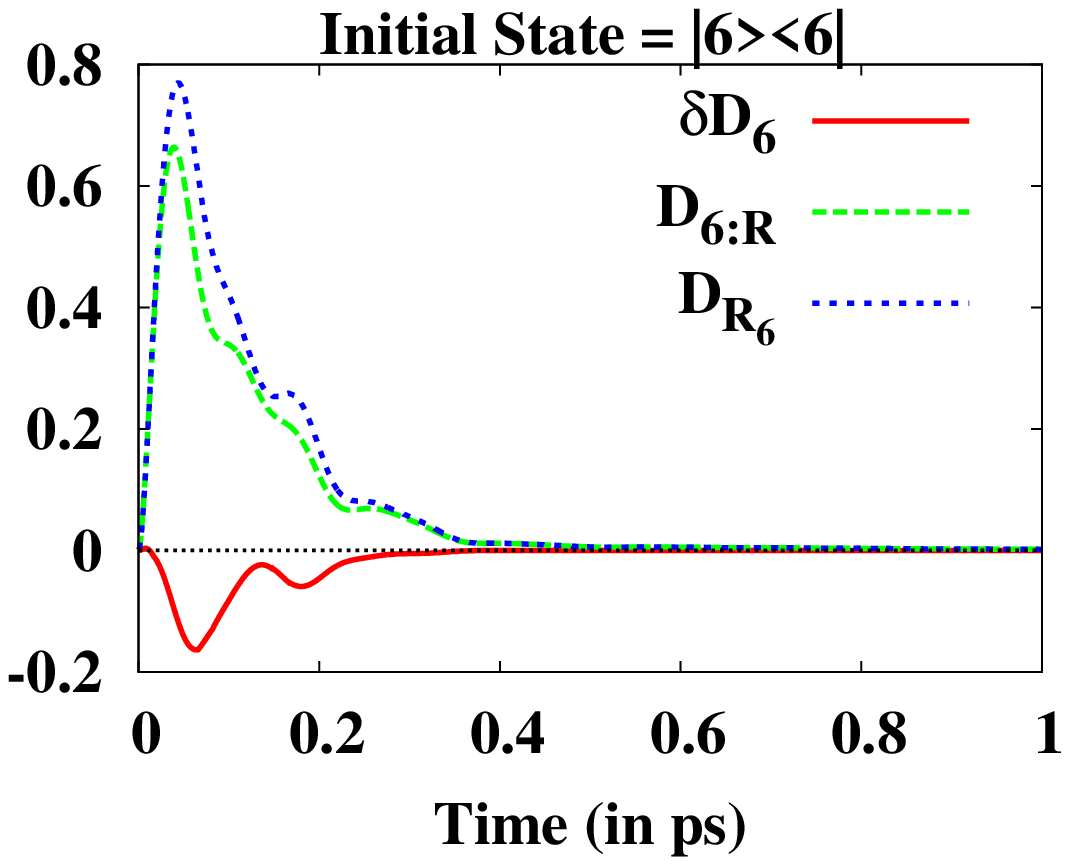}
\includegraphics[width=0.3\linewidth]{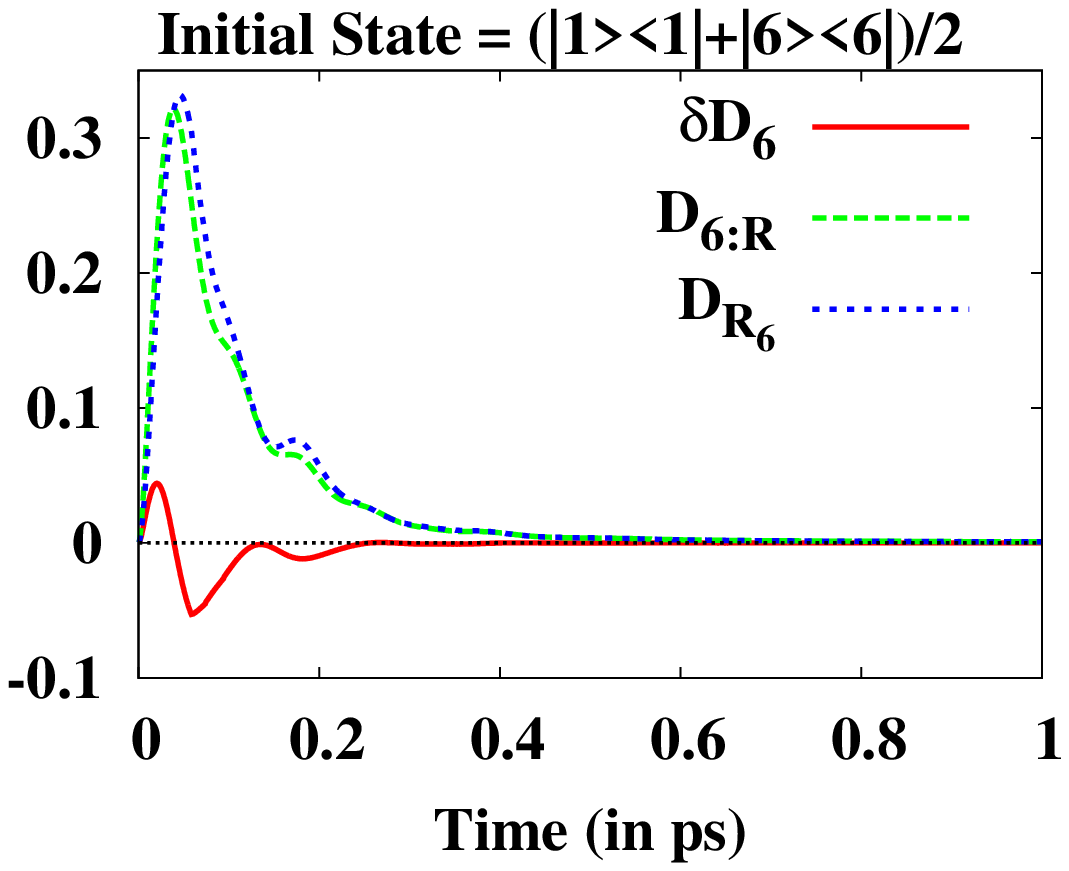}

\caption{(Color online.) Monogamy scores in the FMO complex with the site $5$ as the nodal observer in the first and second rows, and with the site $6$ as the nodal observer in the third and fourth rows. The remaining descriptions are the same as in Fig. \ref{fig:fmo1}.}
\label{fig:fmo3}
\end{figure*}

\subsection{FMO complex}
\label{sec:mngfmo}
Having discussed a relatively simpler model of exciton transportation in light-harvesting complexes, we are now in a position to discuss the distribution of quantum correlations among  different sites in the FMO complex. The system consists of $7$ sites and is governed by Eq. (\ref{eq:evolvedm}) with the Hamiltonian given in Eq. (\ref{FMO_hamil}). The initial state of the evolution is chosen to be among  $\Ket{1}\Bra{1}, \ \Ket{6}\Bra{6}$, and $(\Ket{1}\Bra{1} + \Ket{6}\Bra{6})/2$, as the sites $1$ and $6$ are closer to the receiver (antenna) \cite{SLloyd}. It is to be noted here that the Hamiltonian contains interaction terms between any site, say the $i^{th}$ site, with all the other sites different from the $i^{th}$ site. This may lead to non-zero bipartite quantum correlations among distant sites of the FMO complex. Quantities like $N_{i:R}$, $N_{R_i} = \sum_{j=1,j\neq i}^{N} N(\rho_{j:i})$, and monogamy score for negativity are important to analyze the underlying dynamics of entanglement in the system (similarly for quantum discord). For example, if $N_{i:R}$ is greater than $N_{R_i}$, the multipartite entanglement are more prominent in the system during the dynamics. Otherwise, it is the bipartite contributions that dominate the dynamics. We will show that the dynamics of quantum correlations helps us to indicate the possible structure of the FMO complex.

\begin{figure*}
\includegraphics[width=0.32\linewidth]{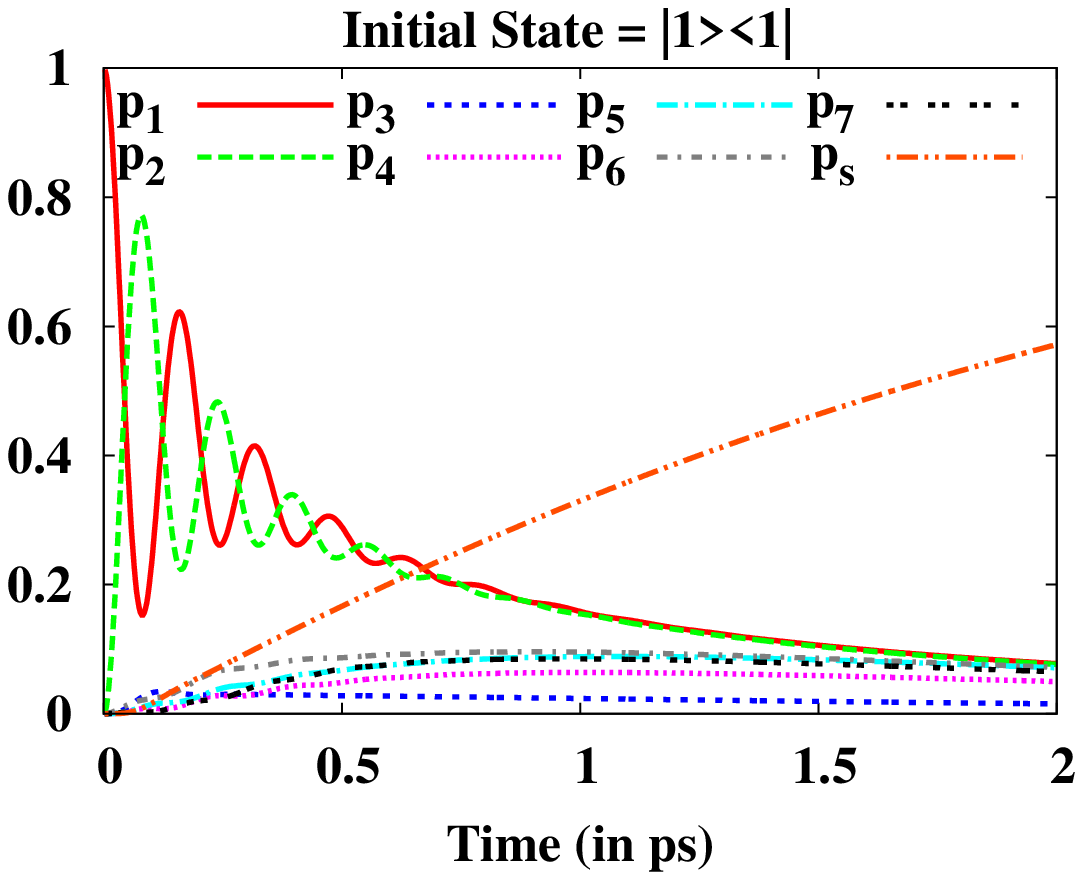}
\includegraphics[width=0.32\linewidth]{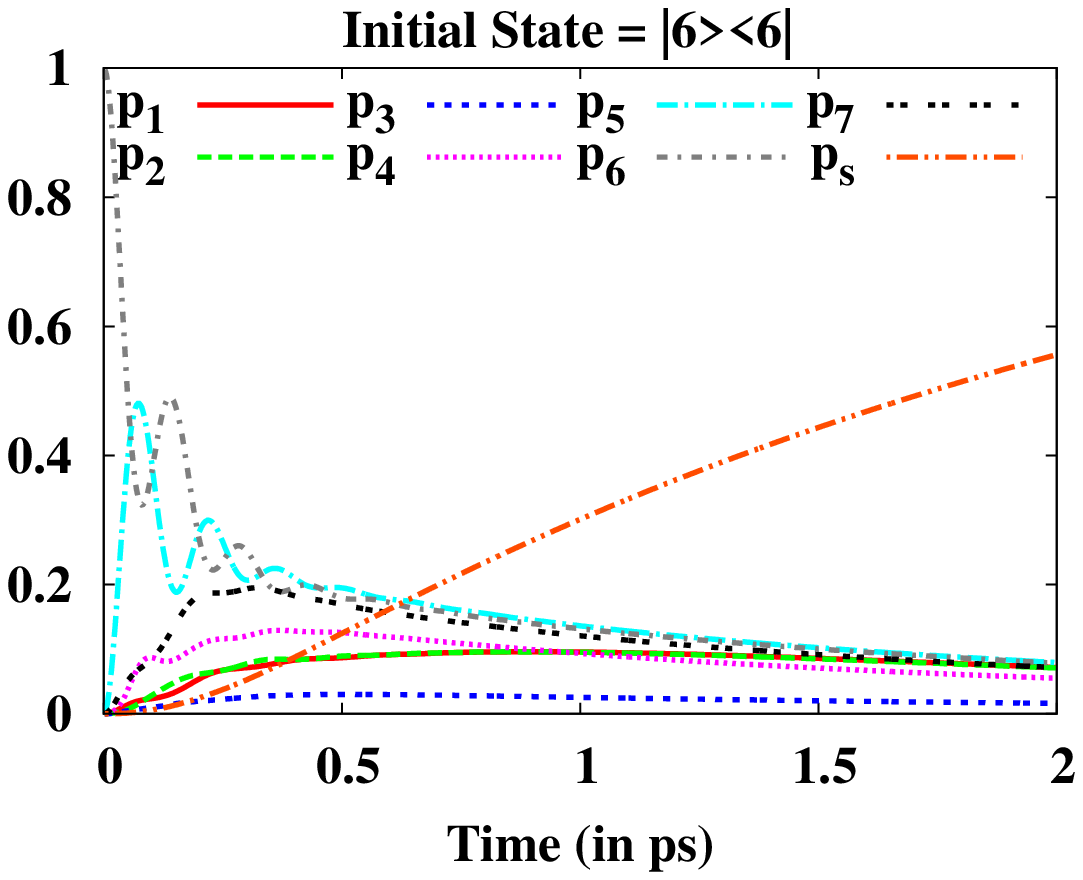}
\includegraphics[width=0.32\linewidth]{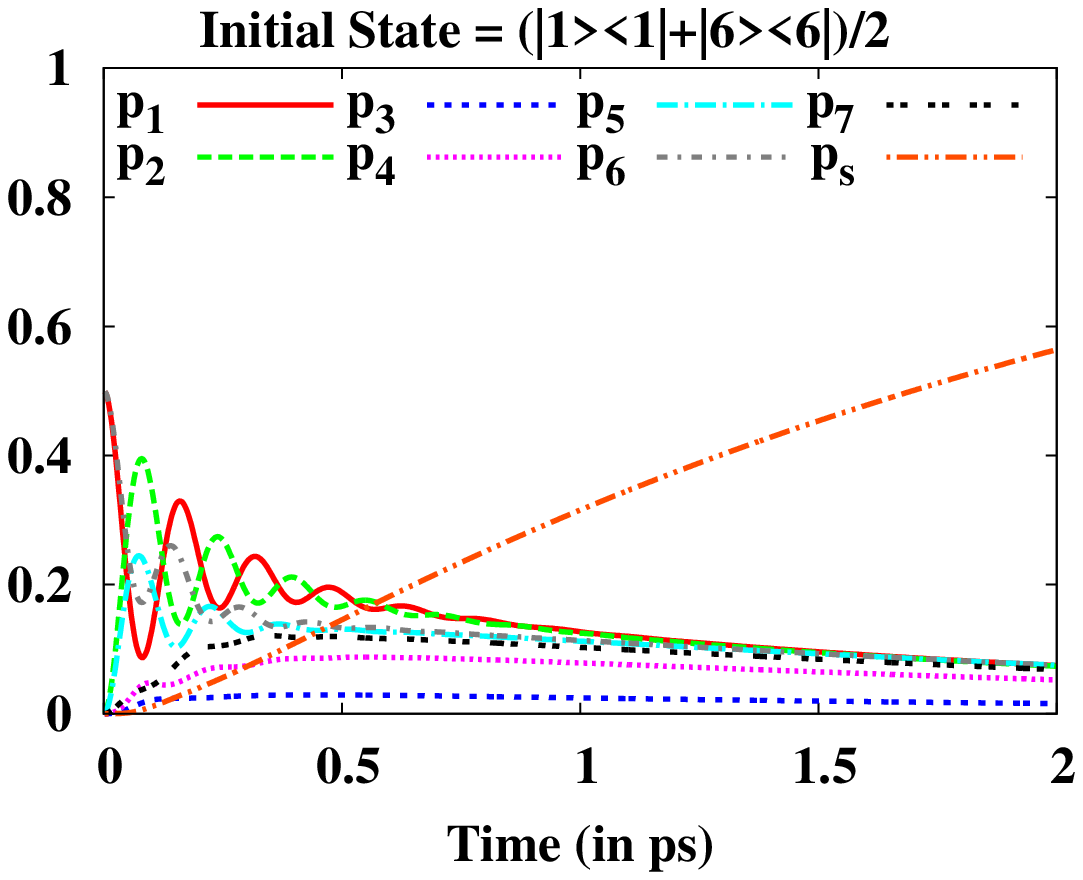}

\caption{(Color online.) Population of different sites of the FMO complex  with time. The population at the $i^{th}$ site is denoted as $p_{i}$, for $i = 1,\ldots,7$, while that of the sink is denoted as $p_{s}$. The vertical axes represent dimensionless parameters. }
\label{fig:popul}
\end{figure*}

\subsubsection{Dynamics of monogamy scores with sites 1 and 2 as nodal observers}

We present here our results regarding  monogamy scores calculated with sites $1$ and $2$, respectively, as nodal observers.  They are close to each other in the on-site energy scale ($E_1 = 215 \ \textrm{cm}^{-1}$ and $E_{2} = 220 \ \textrm{cm}^{-1}$).
We observe that for $\ket{6}\bra{6}$ and $(\arrowvert 1\rangle\langle 1\arrowvert+\arrowvert 6\rangle\langle 6\arrowvert)/2$ as the initial states and the sites $1$ and $2$ as nodal observers, the monogamy score $\delta N_i$ is greater than the bipartite contribution $N_{R_{i}} (i=1,2)$ for most of the time (see  in Fig. \ref{fig:fmo1} for these sites). Note that with these initial states, some small fraction of the population remains in the sites $1$ and $2$ during the dynamics (see Fig. \ref{fig:popul}). The main contribution to $N_{R_i}$ for $i=1,2$ comes from sites $6$ and $5$, i.e., $N_{R_{i}} \approx N_{i:6} + N_{i:5}$ for $i=1,2$. For $\arrowvert 1\rangle\langle 1\arrowvert$ as initial state, $\delta N_i < N_{R_{i}}$ during the initial period in the dynamics, so that we have a non-monogamous nature of the negativity monogamy score there (see in Fig. \ref{fig:fmo1}). It is also observed that after $t > 0.2$ ps, $\delta N_i > N_{R_{i}}$ for the same initial state (see  Fig. \ref{fig:fmo1}). This has not been observed in the cases when initial states were $\arrowvert 6\rangle$ and $(\arrowvert 1\rangle\langle 1\arrowvert+\arrowvert 6\rangle\langle 6\arrowvert)/2$. Specifically, for these initial states,  $\delta N_i > N_{R_{i}}$ for all time.  Note again, and in contrast to the cases when $\arrowvert 6 \rangle$ and $(\arrowvert 1\rangle\langle 1\arrowvert+\arrowvert 6\rangle\langle 6\arrowvert)/2$ were initial states, having $\arrowvert 1\rangle\langle 1\arrowvert$ as the initial state, most of the population remains in the sites $1$ and $2$ during the initial period of time, developing much bipartite entanglement for those sites with other sites in that period. In case of quantum discord, most of the time, $\delta D_{i}$ is negative (non-monogamy), except when the initial state is $(\arrowvert 1\rangle\langle 1\arrowvert+\arrowvert 6\rangle\langle 6\arrowvert)/2$ in which case, quantum discord is monogamous initially and then $\delta D_{i}$ oscillates between positive and negative values. Here, $\delta D_{i}$ degrades quickly in comparison to $\delta N_{i}$, while $N_{R_{i}}$ degrades faster as compared to $D_{R_{i}}$.

\begin{figure}
\includegraphics[scale=0.34]{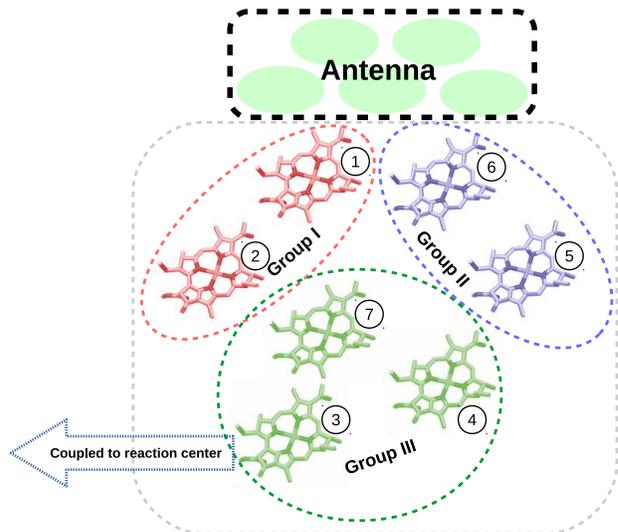}
\caption{(Color online.) Schematic structure of the FMO complex and the group classifications of different sites, as inferred from the dynamics of quantum correlations.}
\label{fig:schematic}
\end{figure}

\begin{figure*}
\includegraphics[scale=0.45]{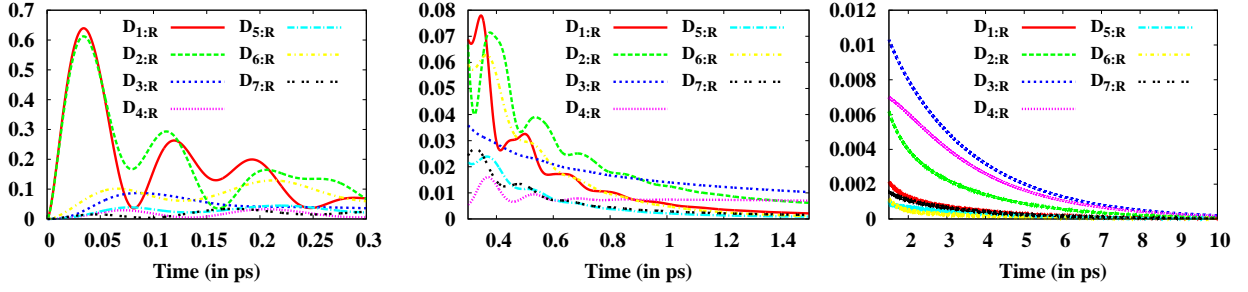}
\caption{(Color online.) Detecting the energy transfer route in the FMO complex when the initial excitation is at the site 1. At any time, the bipartition collection \(\{D_{i:R}\}\) for \(i=1,2, \ldots, 7\) is plotted on the vertical axes. The three panels are for different time spans with different scales on the vertical axes. There are different \(D_{i:R}\) (for different \(i\)) that swims above the others for different times. The energy transfer route is indicated as from site \(j\) to site \(k\) in a certain
time zone, if there is a change, in that time zone, in the site for which the maximum \(D_{i:R}\) occurs, from \(i=j\) to \(i=k\). The vertical axes are measured in bits.}
\label{route1}
\end{figure*}

\begin{figure*}
\includegraphics[scale=0.45]{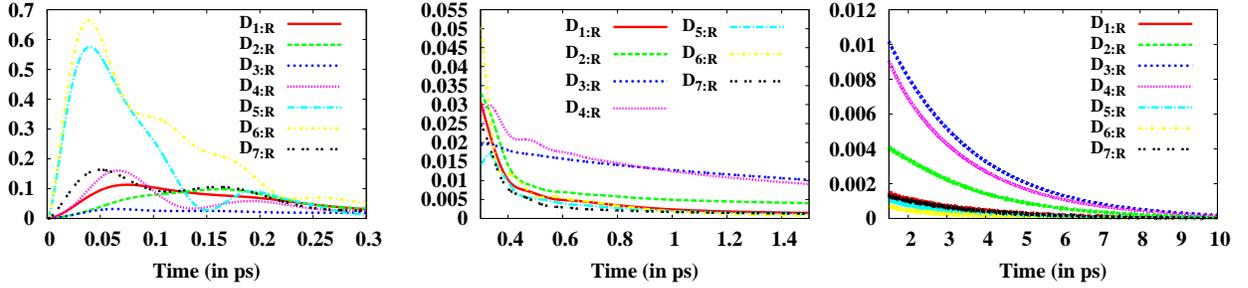}
\caption{(Color online.) Detecting the energy transfer route in the FMO complex when the initial excitation is at the site 6. All other descriptions remain the same as in Fig. \ref{route1}.}
\label{route2}
\end{figure*}

\begin{figure*}
\includegraphics[scale=0.45]{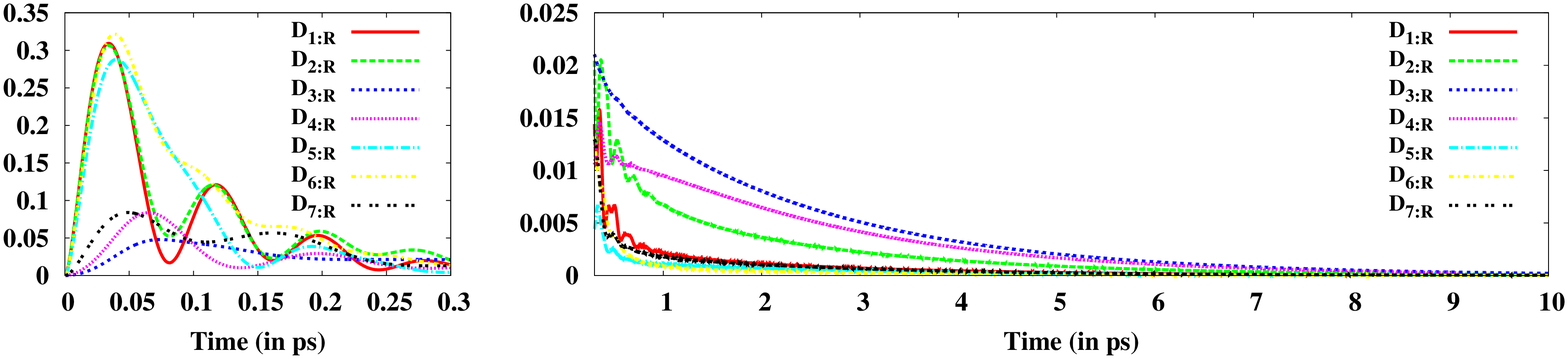}
\caption{(Color online.) Detecting the energy transfer route in the FMO complex when the initial excitation is a mixture of excitations at sites 1 and 6. All other descriptions remain the same as in Fig. \ref{route1}.}
\label{route3}
\end{figure*}

\subsubsection{When sites 3, 4, and 7 are nodal observers}
\label{subsub:347}
Substantial similarity in behavior is obtained with respect to multipartite quantum correlations as measured by monogamy scores with $3$, $4$, and $7$ as nodal observers. For these sites, $\delta N_{i} > N_{R_{i}} (i=3,4,7)$ for most of the time and for any choice of initial state among  $\arrowvert 1\rangle\langle 1\arrowvert$, $\arrowvert 6\rangle\langle 6\arrowvert$, and $(\arrowvert 1\rangle\langle 1\arrowvert+\arrowvert 6\rangle\langle 6\arrowvert)/2$ (see Fig. \ref{fig:fmo2}). These sites are not very close to the antenna and therefore despite a relatively long period of time, only a small fraction of total population ($\approx 10\%$) arrives at these sites (Fig. \ref{fig:popul}). There is correspondingly no substantial amount of bipartite entanglement sustained between any of these sites with other sites. $N_{R_{i}}$ always remains at a lower value as compared to $N_{i:R}$, for $i = 3,4,7$. The discord monogamy score, $\delta D_{i}$, on the other hand shows opposite behavior as compared to negativity monogamy score corresponding to these sites. Precisely, we find that $D_{R_{i}} > D_{i:R}$ for $i=3,4,7$ for any choice of the said initial states. Just like for the cases when the sites $1$ and $2$ are nodal observers, $\delta D_{i}$ degrades quickly as compared to $\delta N_{i}$, while $D_{R_{i}}$ shows more robustness as compared to $N_{R_{i}}$.

\subsubsection{When sites 5 and 6 are nodal observers}

Monogamy scores with sites $5$ and $6$ as nodal observers show qualitative similarity in the dynamics, while being opposite to that for sites $1$ and $2$. However, significant peculiarities are obtained (e.g. in first row middle, third row middle, and fourth row right of Fig. \ref{fig:fmo3}) that necessiates the separation of the  measures considered here with those in Sec. \ref{subsub:347}. The sites $5$ and $6$ are close to each other in the on-site energy scale ($E_5 = 450 \ \textrm{cm}^{-1}$ and $E_{6} = 330  \ \textrm{cm}^{-1}$). One of the peculiarities is that for $\arrowvert 6\rangle\langle 6\arrowvert$ as the initial state, the negativity monogamy $\delta N_{i}  (i=5,6)$ has a lower value as compared to the bipartite term $N_{R_{i}}$ during some initial period of time ($\approx 1.4$ ps). Note that initially, the sites $6$ and $5$ receive maximum fraction of the population (Fig. \ref{fig:popul}) and share a maximum amount of bipartite entanglement with the other sites. It is also noted that when $\arrowvert 6\rangle\langle 6\arrowvert$ is used as the initial state, $\delta N_6$ initially shows non-monogamous nature. For $\arrowvert 1\rangle\langle 1\arrowvert$ and  $(\arrowvert 1\rangle\langle 1\arrowvert+\arrowvert 6\rangle\langle 6\arrowvert)/2$ as initial states, $\delta N_{i} > N_{R_{i}}$ for most of the time interval. Discord monogamy scores ($\delta D_{i}$) remain negative almost all the time in this case. In this case too, $\delta D_{i}$ decays quickly as compared to $\delta N_{i}$, while $D_{R_{i}}$ lasts longer as compared to $N_{R_{i}}$.

\subsubsection{Classification of chromophore sites and structural geometry of FMO complex}
The above discussions on the dynamics of multipartite as well as bipartite quantum correlations ($\delta \mathcal{Q}_i, \ \mathcal{Q}_{i:R}$ and $\mathcal{Q}_{R_i}$) with various sites have enabled us to classify the seven sites into three distinct groups, namely Group I, consisting of sites $1$ and $2$, Group II, with sites $5$ and $6$, and Group III, with sites $3$, $4$, and $7$. The qualitative behaviors of the dynamics of quantum correlations (multiparty as well as bipartite ones) for different sites within  each group are the same, reflecting the fact that different sites within the same group behave similarly in the dynamics of the FMO complex. Recent studies  predict that sites $1$ and $6$ are the closest ones to the chlorosome antenna \cite{SLloyd} and site $3$ is coupled to the reaction center \cite{Renger_biop}. This fact, obtained in Refs. \cite{Renger_biop,SLloyd}, along with our observation on the dynamics of multipartite as well as bipartite quantum correlations strongly suggest the structural arrangement of the FMO complex. Fig. \ref{fig:schematic} shows the schematic structure of the FMO complex, highlighting the group classifications, based on these observations, which match with the results obtained from electron-microscopic studies \cite{Renger_biop,exp_struct}. We will show that these group classifications also support the  primary energy transfer pathway in the FMO complex, obtained in the next section.

\section{Detection of energy transfer route in FMO complex}
\label{sec:energytransfer}

In the preceding section, we focused on the monogamy scores and their constituent expressions to investigate the multiparty quantum correlations in the fully-connected and FMO networks and classified the chromophore sites into three groups. In this section, we will consider collections of quantum correlations in bipartitions for the same purpose. More precisely, we consider the FMO complex and look at the time-evolution of the collections $\{N_{i:R}\}$ and $\{D_{i:R}\}$ as functions of time. We will thereby demonstrate how the dynamics of collections can detect the most probable excitation transfer pathways in the FMO complex. Fig. \ref{route1} shows the dynamics of the quantum discords in the bipartitions $i:R$ for $i=1,2,...,7$ when the initial excitation is at site 1. Upto $0.4$ ps, the maximum between the $D_{i:R}$ at a particular time oscillates between $D_{1:R}$ and $D_{2:R}$, starting with $D_{1:R}$ for $t=0$ ps. From $0.4$ ps to around $0.8$ ps, $D_{2:R}$ becomes the maximum and after that $D_{3:R}$ swims up. Beyond that, $D_{4:R}$ comes close to $D_{3:R}$ and both decay to zero at large time. Such analysis tempted us to infer that the main energy transfer pathway in the FMO complex, when the site 1 was initially excited, is $1 \leftrightarrow 2\leftrightarrow 3\leftrightarrow 4$, i.e., from Group I to Group III (Fig. \ref{fig:schematic}), which is in consistent with earlier findings \cite{chin_njp_2010,Naturephyswhaley,Love}.

Similarly, Fig. \ref{route2} shows the dynamics of collections of quantum discords in bipartition when the initial excitation is at site $6$. In a similar fashion, as when the initial excitation was at site $1$, the dynamics of $\{D_{i:R}\}$,  can detect the energy transfer pathway, and which  in this case is $6 \leftrightarrow 5 \leftrightarrow 4 \leftrightarrow 3$ (i.e., Group II to Group III (Fig. \ref{fig:schematic})). Similar results can also be found with negativity $\{N_{i:R}\}$ and the monogamy score of negativity squared $\{\delta N^2_{i}\}$. For the initial state $(\Ket{1}\Bra{1} + \Ket{6}\Bra{6})/2$, it is found that both the routes, mentioned earlier, come into active consideration (Fig.  \ref{route3}).

\section{Conclusion}
\label{sec:conclusion}
We have presented an analysis of the dynamics of multipartite quantum correlations in light-harvesting complexes modelled by the fully connected and the FMO networks. Several interesting features are shown to be present in the scenario of bipartite vs. multipartite correlations in both the networks. It is been found that in general, multiparty correlations are more prominent and sustain longer than the bipartite ones for entanglement, while the  opposite is found to hold in the case of quantum discord. The discord monogamy score is negative most of the time irrespective of the nodal observer and the initial state indicating that the quantum state of the FMO complex is similar in behavior to the $W$ state \cite{W-state}. Another important feature is that the multipartite monogamy score for quantum discord decays faster than that for negativity, whereas the opposite happens for the bipartite contributions therein. Based on the dynamics of bipartite and multipartite correlations, we have categorized the seven chromophore sites into three distinct groups, which enabled us to predict the structural arrangement of different sites in the FMO complex.  Finally, we have shown that the dynamics of multipartite quantum correlations, as quantified by collections of quantum correlation in bipartition as well as monogamy scores can detect the primary energy transfer pathways in the FMO complex.

\section{Acknowledgement}
We acknowledge computations
performed at the cluster computing facility in
Harish-Chandra Research Institute.

\end{document}